\documentclass{aa}  

\usepackage{graphicx}
\usepackage{txfonts}
\usepackage{hyperref} 
\usepackage{color,soul} 
\usepackage[svgnames]{xcolor}
\usepackage{ulem}
\usepackage{ragged2e} 
\usepackage{arydshln} 
\usepackage{multirow} 

\begin{document} 

   

   \title{Joint inference of the Milky Way star formation history and IMF from {\it Gaia} all-sky $G < 13$ data}

   \titlerunning{Joint SFH and IMF inference from {\it Gaia} all-sky $G < 13$}
   \authorrunning{M. del Alcázar-Julià et al.}

   \author{M. del Alcázar-Julià\inst{1, 2, 3, 4}, F. Figueras\inst{1, 2, 3}, A. C. Robin\inst{5}, O. Bienaymé\inst{6}, F. Anders\inst{1, 2, 3}
          }

    \institute{Departament de Física Quàntica i Astrofísica (FQA), Universitat de Barcelona (UB), C Martí i Franquès, 1, 08028 Barcelona, Spain
         \and
         Institut de Ciències del Cosmos (ICCUB), Universitat de Barcelona (UB), C Martí i Franquès, 1, 08028 Barcelona, Spain 
         \and
         Institut d’Estudis Espacials de Catalunya (IEEC), Edifici RDIT, Campus UPC, 08860 Castelldefels (Barcelona), Spain
         \and
         Isaac Newton Group of Telescopes, Apartado 321, E-38700 Santa Cruz de la Palma, Spain
         \and
         University Marie and Louis Pasteur, OSU THETA, CNRS, Institut UTINAM (UMR 6213), équipe Astro, F-25000 Besançon, France
         \and
         Observatoire astronomique de Strasbourg, Université de Strasbourg, CNRS, 11 rue de l’Université, F-67000 Strasbourg, France}

   \date{Received December 24, 2024; accepted March 10, 2025}

 
  \abstract
   {Despite the fundamental importance of the star formation history (SFH) and the initial mass function (IMF) in the description of the Milky Way, their consistent and robust derivation is still elusive. Recent and accurate astrometry and photometry collected by the {\it Gaia} satellite provide the natural framework to consolidate these ingredients in our local Galactic environment.}
   {We aim to simultaneously infer the IMF and the SFH of the Galactic disc comparing {\it Gaia} data with the mock catalog resulting from the Besançon population synthesis model (BGM). Our goal is also to estimate the impact of the systematics present in current stellar evolutionary models (SEMs) on this inference.} 
   {We use a new implementation of the BGM Fast Approximate Simulations (BGM FASt) framework to fit the seven million star {\it Gaia} DR3 all-sky $G<13$ color-magnitude diagram (CMD) to the most updated dynamically self-consistent BGM.} 
   {Our derived SFH supports an abrupt decrease of the star formation approximately 1-1.5 Gyr ago followed by a significant enhancement with a wide plateau in the range 2-6 Gyr ago. A remarkable hiatus appears around 5-7 Gyr ago with a $\sim$1 Gyr shift depending on the set of stellar models. A complex evolution at ages older than 8 Gyr deserves further investigation. Precise but discrepant values using different SEMs are found for the power-law indices of the IMF. In our fiducial execution with PARSEC SEM, the slope takes a value of $\alpha_2 = 1.45^{+0.19}_{-0.12}$ for the range [0.5-1.53]$M_\odot$, while for masses larger than 1.53 $M_\odot$ we obtain $\alpha_3 = 1.98^{+0.13}_{-0.05}$. Using STAREVOL SEM, the inferred values are $\alpha_2 = 2.48^{+0.09}_{-0.11}$ and $\alpha_3 = 1.64^{+0.15}_{-0.02}$. We find the solution with PARSEC to have a significantly higher likelihood than that obtained with STAREVOL.} 
   {The current implementation of the BGM FASt framework is ready to address executions fitting all-sky {\it Gaia} data up to 14-17 apparent limiting magnitude. This will naturally allow us to derive both a reliable SFH for the early epochs of the Galactic disc evolution and a precise slope for the IMF at low masses.} 


   \maketitle
%

\section{Introduction} \label{sect:introduction}

The star formation history (SFH) of the Milky Way disc is one of the fundamental parameters that characterize the evolution of our galaxy, accounting for the internal and external processes responsible for the variation of the star formation rate (SFR). Entangled with the SFH \citep{Haywood1997, Aumer2009, Kroupa2021}, the initial mass function (IMF; \citealt{Salpeter1955}) plays a crucial role in the chemical evolution of the Milky Way (and stellar systems in general; e.g. \citealt{Burbidge1957, Tinsley1980, Pagel2009}). Therefore, since the middle of the 20th century, the Galactic research community has focused enormous efforts on the precise determination of both the SFH (e.g. \citealt{Cignoni2007, Snaith2015, Bernard2018, Haywood2018, Ruiz-Lara2020, Sysoliatina2021, Cukanovaite2023, Mazzi2024, Gallart2024, Spitoni2024}) and the IMF (e.g. \citealt{Salpeter1955, Schmidt1963, Scalo1998, Kroupa2001, Chabrier2003, Sollima2019, Haghi2020, Li2023, Dickson2023, Kirkpatrick2024}).  

A wide range of tools and techniques have been implemented to determine these fundamental stellar-population parameters of the Galaxy. Most of them rely in some way on star counts as a function of magnitude (e.g. \citealt{Bahcall1980}) or as a function of magnitude and colour \citep{Robin2003, Girardi2005}. With the advent of {\it Hipparcos} \citep{ESA1997, Perryman1995} and {\it Gaia} \citep{GaiaCollaboration2016}, an additional valuable piece of information could be added: their parallax measurements allowed for the study of absolute rather than apparent colour-magnitude-diagrams (CMDs; e.g. \citealt{Vergely2002}) and to construct volume-limited samples (e.g. \citealt{Kirkpatrick2024}). However, none of the previous studies have simultaneously derived both the SFH and the IMF. 

Within the different Galactic models, the Besançon Galaxy Model (BGM; \citealt{Robin2003}) is a holistic population synthesis approach that has been widely used in the last decades for the statistical study of the formation, evolution, and the present content of the Milky Way. From its new strategy \citep{Czekaj2014}, it became possible to directly use the IMF and the SFH to generate a full-sky mock catalog to be compared with data from {\it Tycho-2}, constituting the starting point of the Galactic parameters inference with BGM. Later, \cite{Lagarde2017} presented a new version of BGM incorporating the STAREVOL stellar evolutionary tracks.

At that point, the execution of a full-sky catalog of simulated stars using the standard BGM (hereafter BGM Std) implied a computational cost that could not be assumed for the inference of Galactic parameters using iterative methods. This process demanded the use of statistically robust and physically adequate approximations. To respond to these requirements, \citet{Mor2018} presented the Besançon Galaxy Model Fast Approximate Simulations (BGM FASt), a computationally cheap approximation to BGM that lets us compute fast BGM pseudo-simulations weighting the stars of a BGM Std simulation. Using this new tool and the Approximate Bayesian Computation (ABC) technique, \citet{Mor2019} made a first derivation of the IMF and the SFH of the Galactic thin disc (among other parameters) using {\it Gaia} DR2 data up to $G=12$. 

In parallel, the astrometry outcomes of the {\it Gaia} second data release \citep{GaiaCollaboration2018} opened the window for the improvement of the robustness of the BGM modeling. The problem of dynamical self-consistency was tackled in \cite{Robin2022} by fitting the gravitational potential of the Milky Way to the stellar kinematics and densities of (mostly Solar-neighbourhood) {\it Gaia} DR2 stars down to magnitude $G=17$. At that point, it became clear that it is necessary to re-determine the SFH and IMF of the Milky Way in the Solar neighbourhood. 

We present in this work a new implementation of BGM FASt and, for the first time, its execution considering the full-sky {\it Gaia} DR3 data up to magnitude $G=13$. The high computational demands of the BGM FASt code have now been overcome. The optimized code presented in this work enables future executions at fainter limiting magnitudes.

This article is structured as follows. We present in Sect. \ref{sect:bgm_fast_framework} a basic background on the BGM FASt framework and its capabilities, which is needed to understand the upgrades in the method presented in Sect. \ref{sect:new_bgm_fast}. Then, in Sect. \ref{sect:bgm_fast_g13_results} we show in detail the results of this new implementation, which are contextualized and related to the literature in Sect. \ref{sect:discussion}. Finally, we present the conclusions of this work and some future steps to take in Sect. \ref{sect:conclusions}.


\section{The BGM FASt framework for Galaxy modeling} \label{sect:bgm_fast_framework}

The BGM FASt framework lets us infer and evaluate Galactic parameters by iteratively fitting our model to observed data. It performs computationally cheap pseudo-simulations (PSs) of the Besan\c{c}on Galaxy Model weighting the stars of a pre-sampled BGM Std simulation (so-called mother simulation and labeled MS; \citealt{Mor2018}). BGM FASt is $\approx 10^5$ times faster than BGM Std, which opens a window for evaluating and deriving Galaxy modeling ingredients using Bayesian tools. For illustration, the BGM FASt framework is summarised in Fig. \ref{fig:all_diagram}, while the details of each of its components are given in the following subsections. Following the scheme illustrated in this figure, a BGM FASt pseudo-simulation requires two main inputs: a mother simulation (MS), detailed in Sect. \ref{subsect:basics_BGMFASt}), and a set of parameters to infer ($\Bar{\theta}_k$), whose practical implementation for the derivation of the IMF and the SFH is explained in Sect. \ref{subsect:parameter_space}.

\subsection{The Besançon Galaxy Model} \label{subsect:basics_BGMFASt}

\begin{figure*}
    \centering
    \includegraphics[height=0.91\textheight]{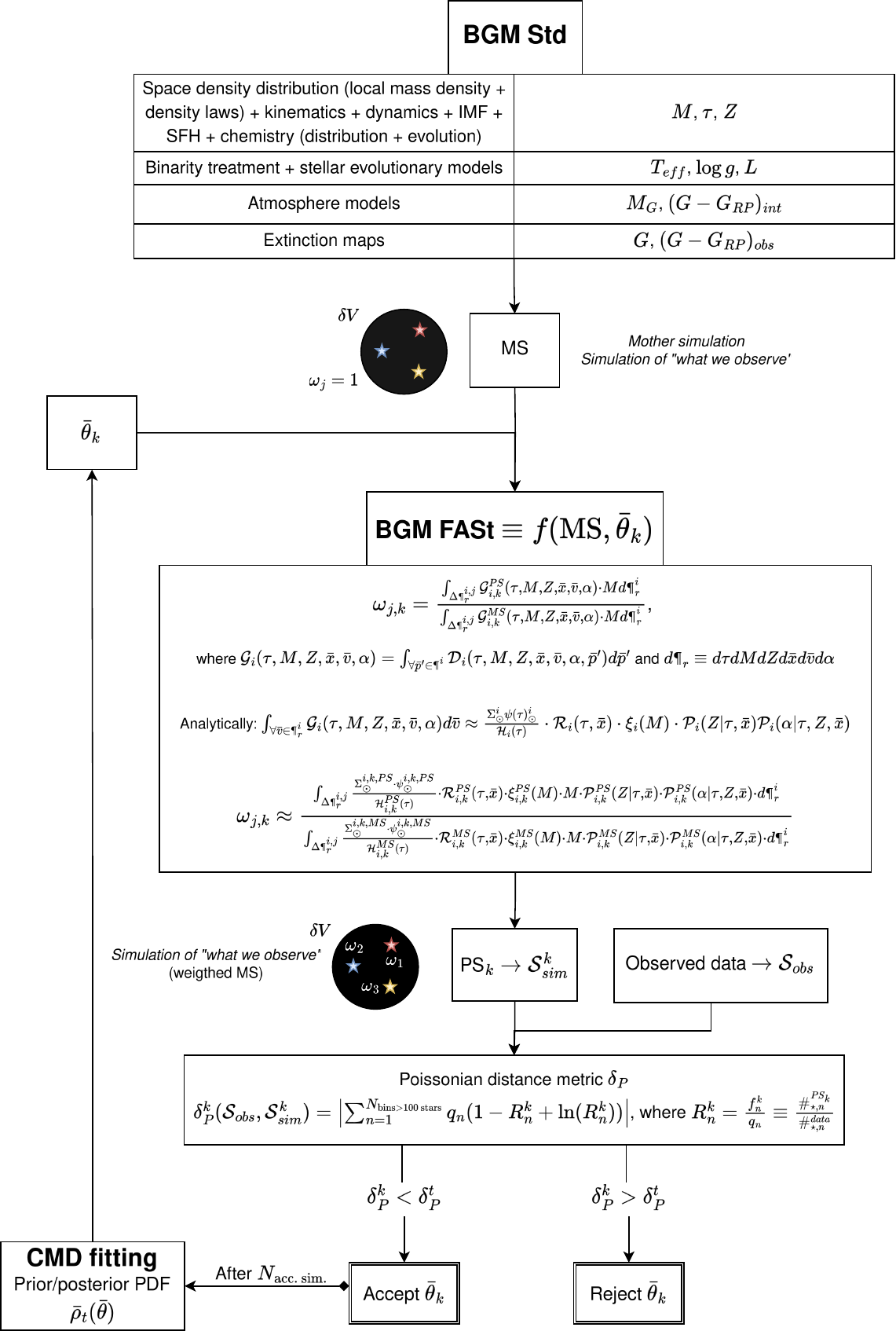}
    \caption{Flux diagram of the process involving BGM Std, BGM FASt, and the CMD fitting technique. BGM Std is summarized in Sect. \ref{subsect:basics_BGMFASt}; we present BGM FASt in Sect. \ref{subsect:bgm_fast}, where the meaning of the subscripts and superscripts $i$ (indicating the $i^{th}$ Galactic component, e.g. thin disc, thick disc, halo, and bar) and $j$ (associated to the $j^{th}$ interval of the reduced parameter space) are explained in detail; regarding the fitting process, see in Sect. \ref{subsect:cmd_technique} the different components of the process as well as an extended explanation of the subscripts and superscripts $k$ (referred to $k^{th}$ iteration within an ABC step), $n$ (indicating the $n^{th}$ bin of the CMD), and $t$ (corresponding to the $t^{th}$ ABC step). See \cite{Mor2018} for a detailed explanation of the equations in the central panel.}
    \label{fig:all_diagram}
\end{figure*}

The mother simulation is the result of applying the population synthesis model BGM Std \citep{Robin2022}. This model operates as follows: first, it uses fundamental functions such as the space density distributions, the IMF, the SFH, and the age-metallicity relation of each Galactic component (e.g. thin disc, thick disc, halo, and bar) to generate stars at a given distance with given masses $M$, ages $\tau$, and metallicities $Z$ in each volume element ($\delta V$ in Fig. \ref{fig:all_diagram}). This process is carried on maintaining the dynamical self-consistency thanks to the last update of BGM Std \citep{Robin2022}, explained below. For binary stars, from the available gas at a given position, BGM Std generates a secondary star following a proportion, binary fraction, semi-major axis, eccentricity, and distribution of the mass-ratio dependent on the mass of the primary \citep{Arenou2011}. 

Then, BGM Std determines the evolutionary stage of each star and, if the age is shorter than the theoretical lifetime, it computes the position of the star in the Hertzsprung–Russell diagram, from which its astrophysical parameters (effective temperature $T_{eff}$, gravity $\log{g}$, and bolometric magnitude $M_{Bol}$) are derived by interpolating in the set of evolutionary tracks. For stars with ages larger than the theoretical lifetime at a given mass, BGM Std computes the mass subsisting as a remnant and the mass released into the interstellar medium, which is instantaneously recycled. 

Finally, as shown in Fig. \ref{fig:all_diagram}, BGM Std converts the astrophysical parameters of the stars into the observed ones following two steps: first, it applies atmosphere models to get the absolute magnitude and the intrinsic colors; and then, from distance and position it includes the effects of reddening of the interstellar medium by means of extinction maps to obtain the apparent magnitude and the observed colors, which constitute the observables that can be compared with the observed catalog data after adapting the magnitudes to the corresponding filters. In addition, BGM Std sets an angular separation limit and merges the different components of the system when they are not resolved. In this work, this value is set to 400 mas to mimic {\it Gaia} DR3 data \citep{Fabricius2021}. 

As said before, the star generation in BGM Std uses fundamental functions that depend on each Galactic component. In the inference process made by BGM FASt (see Sect. \ref{subsect:bgm_fast}), there are two critical global uncertainties of the current best BGM Std model that must be taken into account. On the one hand, the impact of using different stellar evolutionary models on deriving Galactic parameters; on the other hand, how to maintain the dynamical self-consistency of the model with the new set of inferred parameters.

Regarding the choice of stellar models, at present two different stellar evolutionary models are implemented in BGM Std: PARSEC v1.2 \citep{Bressan2012, Chen2014, Chen2015, Fu2018} and STAREVOL \citep{Lagarde2017, Lagarde2019} stellar tracks\footnote{Interpolated PARSEC tracks were obtained from \url{https://github.com/philrosenfield/padova_tracks}, where equally spaced equivalent evolutionary points (EEPs) and an equal number of points between EEPs are provided, which eases the interpolations. STAREVOL tracks have been computed by Nadège Lagarde specifically for the BGM development.}. Both stellar models have their specificities (such as the list of nuclear reaction rates or the equations of state) that the reader can find in the relevant references. A summary of them is presented in Appendix \ref{app:stellar_models}. To compute the {\it Gaia} photometry, we make use of the BTSettl models as provided by \cite{Allard2013} available from IVOA\footnote{\url{https://b2find.eudat.eu/dataset/c29b616f-1dd1-56d1-bece-3876e19bd9a8}}.

The spatial distribution of the stars and their kinematics are set using a dynamical self-consistent solution as described in \cite{Robin2022}. It is achieved by iteratively fitting an axisymmetric potential and the stellar distribution functions of the model to the {\it Gaia} EDR3\footnote{This was first done with {\it Gaia} DR2 and then updated to {\it Gaia} EDR3.} proper motions and parallaxes (see Fig. 2 in \citealt{Robin2022} for a better comprehension of the scheme behind the method and \citealt{Bienayme2018} to understand the physical and mathematical basis of the construction of the distribution functions and potential). Consequently, the dynamical self-consistency of a given simulation is dependent on the input parameters of BGM Std. That is why the modification of BGM Std model ingredients such as the IMF, the SFH, or the stellar evolutionary models will need from a redetermination of the dynamical self-consistency (see Sect. \ref{subsect:parameter_space}).

\subsection{BGM Fast Approximate Simulations} \label{subsect:bgm_fast}

The complexity of the functions in BGM Std, which vary between Galactic components, requires further exploration to improve the model. BGM FASt addresses this need by defining an N-dimensional space of the parameters we want to infer and a tool to perform fast BGM Std simulations to make it possible (see \citealt{Mor2018} for details). We explain here the method for the thin disc, but it can be applied following the same steps to any other Galactic component (e.g. thick disc, halo, and bar). This process is generalized in Fig. \ref{fig:all_diagram}, where the $i^{th}$ Galactic component (in this case, the thin disc) is treated as an additional index. 

We define the parameter space of the thin disc as the $N$-dimensional space containing all the parameters involved in the distribution function of the generated stars of this Galactic component. Mathematically, the parameter space is defined as \textparagraph$ = \tau \times M \times Z \times \Bar{x} \times \Bar{v} \times \alpha \times \Bar{p}'$, where $\tau$, $M$, $Z$, $\Bar{x}$, $\Bar{v}$, and $\alpha$ are the age, the initial mass, the metallicity, the position, the velocity, and the $[\alpha/Fe]$ content of the stars of the thin disc, respectively, and $\Bar{p}'$ accounts for other possible set of independent parameters\footnote{It is mandatory to include $\Bar{p}$' for completeness. At the same time, it opens a window for increasing the complexity of the model in future works.}. From the definition of \textparagraph, we consider a general distribution function of the generated stars $\mathcal{D}(\tau, M, Z, \Bar{x}, \Bar{v}, \alpha, \Bar{p}')$. By marginalizing the general distribution function over $\Bar{p}'$, we obtain the distribution function for the generated thin disc stars in the reduced space \textparagraph$_r$, denoted as $\mathcal{G}(\tau, M, Z, \Bar{x}, \Bar{v}, \alpha)$. 

BGM FASt is based on the idea that the number of stars generated in the $j^{th}$ interval of the reduced parameter space \textparagraph$_r$, $\Delta\text{\textparagraph}_r^j$, is proportional to the mass dedicated to generating stars in that interval \citep{Mor2018}. Therefore, it can be computed as the first moment of the mass: 
\begin{equation} \label{eq:assumption_bgm_fast}
    N_j = K_jM_j = K_j\int_{\Delta \text{\textparagraph}_r^j}\mathcal{G}(\tau, M, Z, \Bar{x}, \Bar{v}, \alpha)\cdot M d\text{\textparagraph}_r,
\end{equation}
where $K_j$ is the particular coefficient of the relation for the $\Delta\text{\textparagraph}_r^j$ interval. Under the assumption that each star corresponds to the smallest possible interval of the parameter space, we associate the $j^{th}$ interval with each individual star. The number of stars in the pseudo-simulation (PS) is then calculated by weighting each star by the proportion between the amount of mass dedicated to generating stars in that interval in the PS and the MS:
\begin{equation} \label{eq:appear_weights}
    \frac{N_j^{PS}}{N_j^{MS}} = \frac{M_j^{PS}}{M_j^{MS}} \Rightarrow N_j^{PS} = \frac{M_j^{PS}}{M_j^{MS}}N_j^{MS} = \omega_jN_j^{MS}, 
\end{equation}
where the weight $\omega_j$ can be expressed in terms of the distribution function as
\begin{equation} \label{eq:compute_weights}
    \omega_j= \frac{\int_{\Delta \text{\textparagraph}_r^j}\mathcal{G}^{PS}(\tau, M, Z, \Bar{x}, \Bar{v}, \alpha)\cdot M d\text{\textparagraph}_r}{\int_{\Delta \text{\textparagraph}_r^j}\mathcal{G}^{MS}(\tau, M, Z, \Bar{x}, \Bar{v}, \alpha)\cdot M d\text{\textparagraph}_r}.
\end{equation}
As previously mentioned, Eq. (\ref{eq:compute_weights}), which forms the core of BGM FASt, can be applied to stars from different Galactic components. In Sect. \ref{subsect:parameter_space} we focus our efforts on applying this equation first to thin disc stars and later to thin and thick disc stars to determine their SFHs and IMF. 

The rigorous mathematical development of the concept as well as the practical implementation shown in Fig. \ref{fig:all_diagram} (how to convert $\mathcal{G}(\tau, M, Z, \Bar{x}, \Bar{v}, \alpha)$ into the magnitudes we want to infer, in this case the SFH and the IMF) are explained in detail in \citet{Mor2018}. Note that in Fig. \ref{fig:all_diagram} there are additional subscripts and superscripts in the magnitudes presented in this section that are associated to the iterative process explained in the next section. 

BGM FASt concludes when the PS is generated by applying the weights from Eq. (\ref{eq:compute_weights}) to the stars in the MS. At this point, the fitting process begins, as shown in Fig. \ref{fig:all_diagram}, using a summary statistics and a distance metric to quantify the similarity between the PS and the observed data.

For the first time, we introduce in this work a new computational strategy that reduces the time required to compute stellar weights by a factor of $\approx 5-7$, significantly improving the efficiency of the inference process compared to previous BGM FASt publications \citep{Mor2018, Mor2019}. 
Taking into account the new implementation of BGM FASt described in Sect. \ref{subsect:parameter_space} and summarized in Eq. (\ref{new_weights}), the weight of a star only depends on its mass and the interval of ages in which it falls. Given the large number of stars in the {\it Gaia} DR3 catalog (over seven million with $G<13$, and a similar number in the MS), many of them will have similar ages and masses. Therefore, we can build a 2-dimensional grid of masses and ages, create an artificial star with the mean properties of mass and age in each cell, and assign the weight of that star to all the stars that fall in the same cell. Following this process we reduce the number of weights to compute from millions to hundreds of thousands. 

Considering steps in mass and age sufficiently small, the influence of this approximation on the final result is negligible. We tested the new method with 100 random combinations of the parameters (IMF and SFH) and found that the mean relative difference between the observed catalog-PS distance (see Sect. \ref{subsect:cmd_technique}) computed with the original version of BGM FASt and the new computationally cheap approximation is below 0.03\%, with the maximum relative difference being of 0.1\%. 

\subsection{BGM FASt fitting strategy} \label{subsect:cmd_technique}

The fitting process is based on fine-tuning the Galactic parameters to make the final PS as similar as possible to the observed catalog data. Thanks to the large number of observed and physical parameters present in the BGM FASt PS, this can be done using different magnitudes from the observable space (e.g. apparent magnitudes, colors, parallaxes...) or derived physical stellar parameters (e.g. masses or ages if we have asteroseismologic observations). Under any of the previous options, there are several elements to define: 1) the summary statistics, 2) the distance metric, and 3) a tool to perform the inference.

\paragraph{Summary statistics.} \label{par:summary_statistics} 
The chosen summary statistics must contain as much information as possible to characterize the parameters we want to infer. When the summary statistics is statistically sufficient to describe a system, then the posterior probability distribution function (PDF) of the parameters under the summarized data is equivalent to the posterior PDF under the full data and the summary statistics is called sufficient statistics \citep{Mor2018}. Since this is practically impossible, we use a summary statistics containing a reduced sample of the data available in the BGM FASt PS and the observed catalog including relevant information for the inference of our parameters. 

We take as a summary statistics a binned modified color-magnitude diagram $M_G'$ vs. {\it Gaia} color (from now on CMD), where $M_G' = G + 5\log_{10}(\varpi/1000) + 5$ is the modified absolute magnitude in which we take into account the broad-band $G$ filter and the {\it Gaia} parallax $\varpi$ (in mas). Note that the computed $M_G'$ differs from the intrinsic $M_G$ because the former is not corrected for the extinction of the interstellar medium. The assumption behind the use of $M_G'$ is that the extinction map in the MS is good enough to characterize the reddening suffered by the light captured by {\it Gaia}. Furthermore, to improve the statistical resemblance between {\it Gaia} and the MS, we affect by errors the astrometric and photometric magnitudes of each star generated with BGM Std using the indications of the {\it Gaia}-DPAC consortium for {\it Gaia} DR3 (see the \href{https://www.cosmos.esa.int/web/gaia/science-performance}{{\it Gaia} instrument model webpage}). 

We want to emphasize that the BGM FASt framework gives us the flexibility to fit any type of data using different summary statistics, from full-sky {\it Gaia} samples up to a given limiting magnitude (as it is done in this work, see Sect. \ref{sect:new_bgm_fast}), to particular regions of the sky or the CMD, always taking into account the role of completeness.

\paragraph{Distance metric.}\label{par:distance_metric} 
Following the initial idea of \citet{Kendall1973} and \citet{Bienayme1987}, already applied in the context of BGM FASt \citep{Mor2017, Mor2018}, we use a Poissonian distance $\delta_P$ to quantify how similar the PS ($\mathcal{S}_{sim}$) and the observed data ($\mathcal{S}_{obs}$) are:
\begin{equation} \label{eq:distance}
    \delta_P(\mathcal{S}_{obs}, \mathcal{S}_{sim}) = \left|\sum_{n=1}^{N_{\text{bins}\geq 100 \text{ stars}}} q_n[1 - R_n + \ln(R_n)]\right|,
\end{equation}
where $R_n = f_n/q_n$ is defined as the quotient between the number of stars in the $n^{th}$ bin of the CMD in the model ($f_n$) and the observed catalog ($q_n$). When the PS fits perfectly the data, $R_n = 1$ $\forall n$ and $\delta_P(\mathcal{S}_{obs}, \mathcal{S}_{sim}) = 0$. Unlike \cite{Mor2018}, we increase here the robustness of the fitting process considering only in the computation of the distance those bins with $\ge 100$ stars in the catalog, which lets us ensure the statistical significance and work with CMDs cleaned from outliers.

\paragraph{Approximate Bayesian Computation.}\label{par:abc}
We use a Python implementation of a sequential Monte Carlo Approximate Bayesian Computation algorithm (SMC-ABC; \citealt{Jennings2017}) to carry out the iterative process responsible for fitting the model to the observed data to estimate the approximate posterior PDF of the parameters ($\Bar{\rho}(\Bar{\theta})$ in Fig. \ref{fig:all_diagram}). The fitting process proceeds as follows. Once we have a PS generated with a given set of the parameters to infer, ABC compares the distance $\delta^k$ between the PS and the observed catalog with the threshold of the $t^{th}$ ABC step. If it is lower, the set of proposed parameters $\Bar{{\theta}}_k$ is accepted and becomes part of the prior PDF of the next step, $\Bar{\rho}_{t+1}(\Bar{\theta})$. Otherwise, it is rejected. 

As shown in Fig. \ref{fig:all_diagram}, the process is carried out in two loops. On the one hand, the distance threshold starts at a determined upper limit that diminishes at each step $t$ until the established lower limit is achieved or the imposed maximum number of steps is reached. On the other hand, at each step the incorporation of sets of parameters $\Bar{\theta}$ to $\Bar{\rho}_{t+1}(\Bar{\theta})$ is repeated until reaching the number of accepted simulations per step set by the user to ensure a statistically robust PDF. While the distance threshold of the step is above the lower limit or the maximum number of steps has not been reached, the derived $\Bar{\rho}_{t+1}(\Bar{\theta})$ becomes part of the prior PDF of the next step. Once the lower limit is achieved or the maximum number of steps reached, we obtain the approximate posterior PDF of the BGM FASt parameters. 

\subsection{Evaluating capabilities of BGM FASt} \label{subsect:add_capabilities}

The parameters inferred with the BGM FASt framework are conditioned to the MS used to derive them and, therefore, to the Galactic ingredients of its receipt. That is why we can consider it as an additional capability of BGM FASt to obtain the approximate posterior PDF of a set of parameters under different combinations of MS ingredients. This lets us evaluate the individual components of the BGM Std modeling. Additionally, the BGM FASt evaluations can be done using particular regions of the CMD more sensitive to the explored Galactic ingredient, or considering alternative summary statistics or distance metrics to the one presented in Eq. (\ref{eq:distance}). See in Sect. \ref{subsect:parameter_space} a practical application of this concept focusing on different stellar evolutionary models.

\section{BGM FASt G13 implementation} \label{sect:new_bgm_fast}
In this section, we present the ingredients used in the current implementation of BGM FASt for the best fitting to the full-sky {\it Gaia} data up to limiting magnitude $G=13$.

\subsection{Mother simulation ingredients} \label{subsect:mss_ingredients}

Most of the MS ingredients used in this work are described in detail in \citet{Robin2022}. We use, for the thin disc, the non-parametric SFH implemented there. In the case of the thick disc SFH, we consider two different possibilities, one flat and another Gaussian centered at 10 Gyr ago and with a standard deviation of 2 Gyr. In both cases, the thick disc SFHs span from 8 to 12 Gyr ago, are truncated outside this range, and their total surface density is the same. The reason why we consider two different shapes for the thick disc SFH is that we aim to run different BGM FASt executions, some fitting the thick disc SFH (for which an MS with a flat thick disc SFH is a better option to avoid influencing the final results), and some others fixing it (in these cases an MS with a Gaussian SFH is more realistic). The values of the SFHs for thin and thick discs in the MSs are found in Table \ref{tab:results}. 

For the IMF, we consider a three-truncated power-law $\xi(M) \propto M^{-\alpha}$ with slopes $(\alpha_1, \alpha_2, \alpha_3) = (1.0, 1.7, 2.4)$ corresponding to the mass ranges shown in Table \ref{tab:results}. The slopes are very similar to the values of ($\alpha_1, \alpha_2, \alpha_3) = (0.85, 1.76, 2.5)$ used in \citet{Robin2022}. Regarding the stellar evolutionary models, as said in Sect. \ref{subsect:basics_BGMFASt}, we use PARSEC and STAREVOL tracks. 


\subsection{Parameter space for IMF and SFH fitting} \label{subsect:parameter_space}

In \citet{Mor2019} we chose a 15-dimensional space including the three slopes of the IMF, nine parameters of a non-parametric SFH covering the range of ages of the thin disc, its radial scale length and the volume stellar mass density of the young and old thick discs. In this new work, two different frameworks are implemented: one in which we consider a 13-dimensional parameter space composed of two IMF slopes ($\alpha_2$ and $\alpha_3$) and 11 parameters of a non-parametric SFH for the thin disc; and another in which the thick disc SFH is unfrozen including in the fit four additional parameters describing it, resulting in a 17-dimensional parameter space. Adding the thick disc SFH parameters in the fit now appears as a possibility thanks to the increase in limiting magnitude. At $G<12$, the thick disc represented $\approx$7\% of the sample, while this number rises to $\approx$10\% at $G<13$. This contribution is expected to grow in future works using fainter magnitudes. On the other hand, at limiting magnitude $G=13$ we cannot fit the low-mass IMF (too few stars), so we fix $\alpha_1=1.0$, in line with the value it has in the MSs. In addition, the slopes derived in this work correspond to a time- and space-averaged IMF, neglecting the possible variations of the IMF with Galactic epochs and from cluster to cluster (e.g. \citealt{Jerabkova2018, DibandBasu2018}), whose modeling and implementation can be included in BGM FASt future executions in a consistent manner.

We assume for the current runs the MSs are dynamically self-consistent (see Sect. \ref{subsect:basics_BGMFASt}), which allows us to exclude the radial scale length from the inference parameter space. The limitation of this process is that we are breaking this dynamical self-consistency throughout the inference process. That is why the whole process requires iterating between BGM Std and BGM FASt to converge into a final solution with the best IMF and SFHs in a dynamically self-consistent context (see Sect. \ref{subsect:limits_and_caveats}). 

See in Table \ref{tab:results} the mass and age ranges of each parameter. Table \ref{tab:summary_executions} shows the set of different configurations considered in this work and named G13(P/S)-(13/17), depending on the evolutionary model that is used to construct the MS ("P" stands for PARSEC and "S" for STAREVOL) and the number of BGM FASt fitted parameters ("13" or "17" if the thick disc SFH is frozen or not, respectively). 

Finally, we center the Gaussian priors on the MSs values of the parameters, that are found in Table \ref{tab:results}). This responds to the fact that we consider the MSs the departing point for BGM FASt and, therefore, the values used for their generation constitute our best knowledge before applying the inference method.

\begin{table}[h!]
    \centering
    \begin{tabular}{c|c|c|c}
         & MS SEM & MS thick disc SFH & Fit thick disc \\ \hline
    G13P-13 & PARSEC & Gaussian & No \\
    G13S-13 & STAREVOL & Gaussian & No \\
    G13P-17 & PARSEC & Flat & Yes \\
    G13S-17 & STAREVOL & Flat & Yes
    \end{tabular}
    \caption{Summary of the characteristic of the BGM FASt executions in the present work including their name, the stellar evolution model (SEM), the thick disc SFH used to generate the mother simulation (MS), and whether we include the thick disc or not in the fitting process. The name of the simulations is built as follows: "G13" (limiting magnitude) + "P" or "S" (SEM) + "-" + "13" or "17" (number of fitted parameters, that is fixing or fitting the four parameters of the thick disc SFH).}
    \label{tab:summary_executions}
\end{table}

\begin{figure*}
    \centering
    \includegraphics[width=\textwidth]{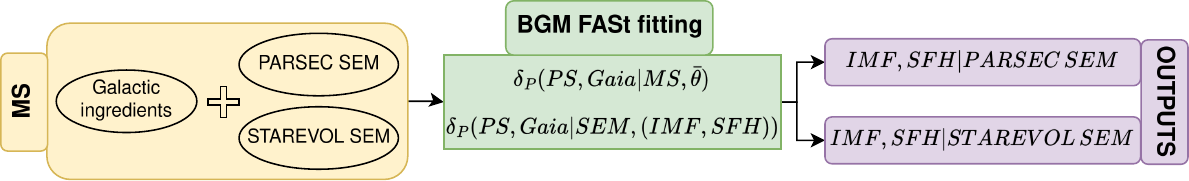}
    \caption{Conditional inference scheme. This figure shows how the output parameters describing the IMF and the SFH are derived conditioned to the ingredients of the MSs and, more particularly, to the two stellar evolutionary models (SEMs) considered in this work. Read Sect. \ref{subsect:parameter_space} for a more detailed description of the process. Besides considering two different SEMs, we also run BGM FASt on MSs built with two different thick disc SFHs. Considering all combinations (two SEMs and two thick disc SFHs), we have in total 4 MSs over which we perform the BGM FASt executions (see Tables \ref{tab:summary_executions} and \ref{tab:results}).}
    \label{fig:conditional_inference}
\end{figure*}

This new implementation of BGM FASt implies some modifications of the original equations \citep{Mor2018}. For instance, the computations regarding the density laws with the Einasto profiles are no longer in use. Under this consideration, we can suppress the spatial dependencies, keeping only the IMF and the SFH as the BGM FASt parameters to infer. The weight equation, therefore, is extremely simplified to
\begin{equation} \label{new_weights}
    \omega_j = \frac{\int_{\Delta\tau_j} \sum_{\odot}^{PS}(\tau)d\tau \int_{\Delta M_j}\xi^{PS}(M)M dM}{\int_{\Delta\tau_j} \sum_{\odot}^{MS}(\tau)d\tau \int_{\Delta M_j}\xi^{MS}(M)M dM},
\end{equation}
where $\sum_{\odot}(\tau)$ is the star formation rate of the thin/thick disc at the Solar neighborhood at time $\tau$, $\xi(M) = \propto M^{-\alpha}$ is the value of the IMF at $M$, we consider $d\tau = \Delta \tau_j$ equal to the age intervals given in Table \ref{tab:results}\footnote{We assume a constant SFH within each age interval.}, and we take $\Delta M_j = [M_j, M_j + 0.025 M_\odot]$, being $M_j$ the mass of the $j^{th}$ star. The value of $\omega_j$ obtained from Eq. (\ref{new_weights}) is then corrected taking into account stellar multiple systems arising from the imposed IMF and SFH (see Sect. 3.3 from \citealt{Mor2018} for more details).

By construction (see Sect. \ref{subsect:add_capabilities}), the derivation of the IMF and SFH parameters in BGM FASt depends on the ingredients used to generate the MSs and not fitted in the process. This fact confers to the BGM FASt framework the great advantage of evaluating them. In this case, we take benefit of this intrinsical feature by running BGM FASt on MSs generated using different stellar evolutionary models (SEMs)\footnote{We also consider different thick disc SFHs, as explained in Sect. \ref{subsect:mss_ingredients}. However, we don't do that to evaluate them, but to make the different MSs suitable to the particular fitting process applied to each of them.}. The comparison of the obtained best fit to {\it Gaia} data allows us to analyze the behavior of each of them. We describe this process in Fig. \ref{fig:conditional_inference}: 1) we consider MSs made up of the same Galactic ingredients except for the stellar evolutionary models, for which we generate an MS with PARSEC and another with STAREVOL; 2) we run the BGM FASt fitting on the MSs. This can be understood as the computation, at each step, of the distance between the PS and {\it Gaia} assuming a given SEM, IMF, and SFH; 3) we obtain as output the approximate posterior PDF of the IMF and SFH parameters conditioned to each of the MSs, therefore, to each of the stellar evolutionary models used to build them. The results of this evaluation can be found in Sect. \ref{subsect:cmds_evolution}. 


\subsection{CMD description} \label{subsect:cmd_description} 

As said before (see Sect. \ref{subsect:cmd_technique}), our summary statistics comprises a 2-dimensional CMD containing $M_G'$ in terms of a {\it Gaia} color. In this case, we consider the CMD fitting within the limits $-5\le M_G'\le 8.5$. We go up to absolute magnitude $M_G'=8.5$ to be able to fit the entire mass range of $\alpha_2$ considering the limiting apparent magnitude $G=13$ of the samples\footnote{This value of $M_G'=8.5$ is derived from considering a PARSEC isochrone of $\log{(age/yr)}=8$ and determining the absolute magnitude at which stars with masses around 0.5 $M_\odot$ are found. At $G<13$, 99\% of stars with $M<0.6 M_\odot$ that can be observed are at a distance $<300$ pc. At this distance, extinction is mostly negligible and $M_G' \approx M_G$.}. Note that this choice differs from the previous implementations of BGM FASt \citep{Mor2018, Mor2019}, in which they fit the CMD in the range $-1\le M_G'<5$ and the luminosity function (1-dimensional) for $5\le M_G'<8.5$. This change is possible because we have doubled the number of stars at $G<13$ with respect to $G<12$ and we ensure now the statistical robustness in the computation of the distance (see Sect. \ref{par:distance_metric}).

For the {\it Gaia} color, we use $G - G_{RP}$ instead of $G_{BP} - G_{RP}$. This choice is supported by the following arguments: 1) $G - G_{RP}$ is less affected by extinction than $G_{BP} -G_{RP}$, which directly implies for the latter the loss of more than 300,000 stars due to the limitations of the photometric transformations of \cite{Evans2018} ($-0.47 \leq G_{BP} - G_{RP} \leq 2.73$) and the lack of photometric observations in the $G_{BP}$ passband for some stars; 2) for the same reason, the {\it Gaia} color $G - G_{RP}$ is less dependent on the selected extinction map; 3) using $G - G_{RP}$ instead of $G_{BP} - G_{RP}$ we are able to consider colder stars in the working area of our CMDs, which is crucial to fit the old SFH and the low-intermediate-mass range of the IMF, and 4) the $G_{BP}$ passband in {\it Gaia} has been redefined several times, and the BTSettl grid used in the last version of BGM Std does not include the last modifications. The only drawback we have found against using $G-G_{RP}$ instead of $G_{BP}-G_{RP}$ is that the former's CMDs appear slightly more concentrated than the latter's (due to the smaller wavelength coverage and the shift towards the blue of the $G_{BP}$ passband), which may imply a dilution of the physical features along the color axis. Counteracting this drawback, the error of the {\it Gaia} color $G_{BP} - G_{RP}$ grows much faster with respect to the error in $G - G_{RP}$ when increasing the apparent magnitude. All in all, taking into account that our aim in the future is to apply our strategy to samples up to fainter limiting, we consider that the advantages of using $G - G_{RP}$ overcome its drawbacks.

Regarding the binning, we consider steps of $\Delta (G-G_{RP}) = 0.05$ mag and $\Delta M_G' = 0.25$ mag, which gives us a reasonable relation between the conservation of the stars' information and a statistically sufficient number of stars per bin. Finally, we divide the fit into three different CMDs characterizing different Galactic latitude ranges: $|b|<10$, $10<|b|<30$, and $|b|>30$. By doing that, we avoid losing valuable information on the effects of a given combination of IMF and SFH in different regions or components (e.g. thin and thick discs, and halo). The catalogue data used for the fitting is the entire {\it Gaia} DR3 $G<13$ sample of stars with positive parallaxes and available $G - G_{RP}$ colors \citep{GaiaCollaboration2022}, sampling inhomogeneously a volume of 1 (55\% of the stars) to 4 (95\% of the stars) kpc.

\subsection{Best ABC parameters} \label{subsect:abc_parameters} 

The parameters to be set by the Approximate Bayesian Computation (ABC) include the distance thresholds, the number of accepted simulations per step, and the maximum number of steps. For the former, the goal is to establish intelligent upper and lower limits to obtain a statistically robust posterior approximate PDF of the parameters representing a scientific improvement with respect to previous studies. That is why we choose for the upper limit $\delta_{max}$ the distance between the MSs and the {\it Gaia} DR3 data. Behind this selection, we accept that the MSs are currently the best modeling of the Milky Way within the BGM Std framework. For the lower limit $\delta_{min}$, we set an arbitrarily small value that lets us explore the possibilities of the inference process without constraints. 

\begin{figure}[!ht]
   \centering
   \includegraphics[width=\hsize]{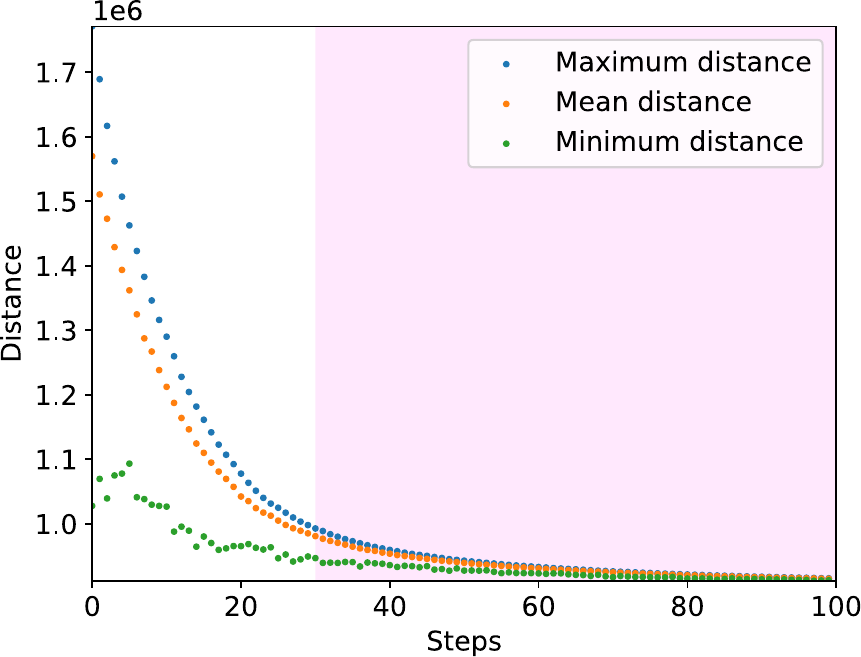}
    \caption{Maximum, minimum, and mean distance evolution over ABC steps for the BGM FASt execution G13P-13. The maximum distance can be considered the threshold of the distance at a given step. The pink-shaded region covers the last 70 steps, which are considered for the computation of the posterior approximate PDFs of the parameters to avoid the priors' influence.}
    \label{fig:distance_evolution}
\end{figure}

For each step, we run as many simulations as needed to collect 500 accepted sets of parameters. This value was decided after checking that the final solution is stable and equivalent within the statistical fluctuations to the one derived by doubling the number of accepted simulations per step. As seen in Fig. \ref{fig:distance_evolution}, we fix the maximum number of steps to 100. In Fig. \ref{fig:parameters_evolution} we show the convergence of each of the fitted parameters. As can be seen, the chosen value of 100 steps is suitable for achieving the convergence in the parameters with enough information (see Appendix \ref{app:evolution_parameters} for details). No better results are observed by increasing the number of steps to 200. For the computation of the final posterior of the parameters, we discard the first 30 steps to avoid the influence of the priors and keep the information of the equivalent solutions (degeneracies).

\section{The {\it Gaia} G13 - BGM FASt scientific outputs} \label{sect:bgm_fast_g13_results}

In Sect. \ref{subsect:imf_and_sfh} we present and analyze the approximate posterior PDFs of the inferred parameters describing the IMF and the SFHs. The best color-magnitude diagrams resulting from the BGM FASt fits are examined and compared in Sect. \ref{subsect:cmds_evolution}, where we evaluate PARSEC and STAREVOL evolutionary models against {\it Gaia} data. 


\begin{table*}[h!]
    \caption{Outputs from the BGM FASt inference for the four runs made up to magnitude $G<13$ ("G13") considering PARSEC or STAREVOL as stellar evolution model ("P" or "S"), and fixing or fitting the thick disc SFH (therefore fitting "13" or "17" parameters). Columns are: parameter name, mass ($M_\odot$) or age (Gyr) range, value in the MS, posterior of the parameter in each of the executions (G13P-13, G13S-13, G13P-17, and G13S-17), and units of each parameter. The fitted parameters are: two IMF slopes, 11 SFH surface densities for the thin disc, and 4 SFH surface densities for the thick disc (for G13P-17 and G13S-17). For convenience, we also show the first slope of the IMF $\alpha_1$ (fixed) and, in parenthesis, the value considered when the parameter is not fitted. In the last row, we show the total surface density resulting from the inference. In the MS column, we separate with a slash the thick disc SFH parameters $\Sigma_\odot^{9T}$, $\Sigma_\odot^{10T}$, $\Sigma_\odot^{11T}$, and $\Sigma_\odot^{12T}$ considered for the executions G13P-13 and G13S-13, and those corresponding to the runs G13P-17 and G13S-17.} \label{tab:results}
    \centering
    \resizebox{0.8\textwidth}{!}{
    \begin{tabular}{c c c c c c c c c}        
    \hline\hline \vspace{-8pt} \\
    \vspace{3pt}
    \textbf{Parameter} &   \textbf{Mass/Age range} & \textbf{MS} & \pmb{$\Bar{\theta}_{\text{G13P-13}}$} & \pmb{$\Bar{\theta}_{\text{G13S-13}}$} & \pmb{$\Bar{\theta}_{\text{G13P-17}}$} & \pmb{$\Bar{\theta}_{\text{G13S-17}}$} & \textbf{Units} \\   
    \hline \hline \vspace{-8pt} \\
    $\alpha_1$ & 0.015-0.5 & 1.0 & (1.0) & (1.0) & (1.0) & (1.0) & Dimensionless \\
    \vspace{1.5pt}
    $\alpha_2$ & 0.5-1.53 & 1.7 & $1.45^{+0.19}_{-0.12}$ & $2.48^{+0.09}_{-0.11}$ & $1.32^{+0.22}_{-0.13}$ & $2.52^{+0.11}_{-0.13}$ & Dimensionless \\
    \vspace{3pt}
    $\alpha_3$ & 1.53-120 & 2.4 & $1.98^{+0.13}_{-0.05}$ & $1.64^{+0.15}_{-0.02}$ & $2.02^{+0.12}_{-0.05}$ & $1.80^{+0.12}_{-0.03}$ & Dimensionless\\
    \hline \vspace{-6.5pt} \\
    \vspace{1.5pt}
    $\Sigma^0_\odot$ & 0-0.15 & 1.6 & $1.10^{+0.16}_{-0.16}$ & $2.26^{+0.18}_{-0.30}$ & $1.04^{+0.19}_{-0.16}$ & $2.06^{+0.27}_{-0.23}$ & $M_\odot$pc$^{-2}$Gyr$^{-1}$ \\
    \vspace{1.5pt}
    $\Sigma^1_\odot$ & 0.15-1 & 2.5 & $0.55^{+0.08}_{-0.11}$ & $1.69^{+0.18}_{-0.33}$ & $0.51^{+0.09}_{-0.10}$ & $1.35^{+0.16}_{-0.23}$ & $M_\odot$pc$^{-2}$Gyr$^{-1}$ \\
    \vspace{1.5pt}
    $\Sigma^2_\odot$ & 1-2 & 2.1 & $2.53^{+0.27}_{-0.45}$ & $7.95^{+0.75}_{-1.73}$ & $2.35^{+0.32}_{-0.36}$ & $6.00^{+0.66}_{-0.99}$ & $M_\odot$pc$^{-2}$Gyr$^{-1}$ \\
    \vspace{1.5pt}
    $\Sigma^3_\odot$ & 2-3 & 1.7 & $3.42^{+0.61}_{-0.73}$ & $7.05^{+1.11}_{-1.80}$ & $3.13^{+0.65}_{-0.65}$ & $5.01^{+1.15}_{-1.22}$ & $M_\odot$pc$^{-2}$Gyr$^{-1}$ \\
    \vspace{1.5pt}
    $\Sigma^4_\odot$ & 3-4 & 1.8 & $4.70^{+1.22}_{-1.69}$ & $7.61^{+1.58}_{-2.31}$ & $4.72^{+1.18}_{-1.79}$ & $5.55^{+1.45}_{-1.78}$ & $M_\odot$pc$^{-2}$Gyr$^{-1}$ \\
    \vspace{1.5pt}
    $\Sigma^5_\odot$ & 4-5 & 1.8 & $4.42^{+1.54}_{-1.96}$ & $5.13^{+1.90}_{-1.98}$ & $3.68^{+1.79}_{-1.63}$ & $4.48^{+1.80}_{-1.76}$ & $M_\odot$pc$^{-2}$Gyr$^{-1}$ \\ 
    \vspace{1.5pt}
    $\Sigma^6_\odot$ & 5-6 & 2.3 & $3.28^{+1.74}_{-1.66}$ & $3.94^{+2.35}_{-1.81}$ & $3.84^{+1.88}_{-1.79}$ & $2.30^{+2.25}_{-1.20}$ & $M_\odot$pc$^{-2}$Gyr$^{-1}$ \\
    \vspace{1.5pt}
    $\Sigma^7_\odot$ & 6-7 & 2.3 & $0.74^{+2.36}_{-0.26}$ & $8.85^{+2.85}_{-3.07}$ & $0.97^{+2.75}_{-0.38}$ & $6.25^{+2.00}_{-2.87}$ & $M_\odot$pc$^{-2}$Gyr$^{-1}$ \\
    \vspace{1.5pt}
    $\Sigma^8_\odot$ & 7-8 & 3.4 & $3.28^{+2.59}_{-1.76}$ & $10.49^{+3.03}_{-3.41}$ & $4.39^{+2.72}_{-2.06}$ & $5.74^{+2.63}_{-2.40}$ & $M_\odot$pc$^{-2}$Gyr$^{-1}$ \\
    \vspace{1.5pt}
    $\Sigma^9_\odot$ & 8-9 & 3.4 & $8.88^{+2.83}_{-3.18}$ & $2.13^{+2.49}_{-1.13}$ & $9.07^{+2.83}_{-3.29}$ & $1.50^{+2.48}_{-0.72}$ & $M_\odot$pc$^{-2}$Gyr$^{-1}$ \\
    \vspace{1.5pt}
    $\Sigma^{10}_\odot$ & 9-10 & 3.4 & $14.08^{+2.25}_{-5.56}$ & $4.16^{+1.97}_{-1.81}$ & $13.09^{+2.60}_{-5.19}$ & $4.14^{+1.66}_{-2.03}$ & $M_\odot$pc$^{-2}$Gyr$^{-1}$ \\ 
    \vspace{1.5pt}
    $\Sigma^{9T}_\odot$ & 8-9 & 1.5 / 1.7 & (1.5) & (1.5) & $1.48^{+1.19}_{-0.87}$ & $14.56^{+0.67}_{-5.34}$ & $M_\odot$pc$^{-2}$Gyr$^{-1}$ \\ 
    \vspace{1.5pt}
    $\Sigma^{10T}_\odot$ & 9-10 & 1.9 / 1.7 & (1.9) & (1.9) & $0.69^{+1.67}_{-0.25}$ & $0.63^{+2.61}_{-0.15}$ & $M_\odot$pc$^{-2}$Gyr$^{-1}$ \\ 
    \vspace{1.5pt}
    $\Sigma^{11T}_\odot$ & 10-11 & 1.9 / 1.7 & (1.9) & (1.9) & $0.85^{+1.54}_{-0.44}$ & $0.81^{+1.40}_{-0.40}$ & $M_\odot$pc$^{-2}$Gyr$^{-1}$ \\ 
    \vspace{8pt}
    $\Sigma^{12T}_\odot$ & 11-12 & 1.5 / 1.7 & (1.5) & (1.5) & $1.48^{+1.48}_{-0.90}$ & $2.68^{+1.02}_{-1.22}$ & $M_\odot$pc$^{-2}$Gyr$^{-1}$ \\ 
    \vspace{3pt}
    $\Sigma^{T}_\odot$ & all & 31.4 & $53.72^{+5.48}_{-6.89}$ & $67.20^{+6.82}_{-5.99}$ & $53.42^{+6.32}_{-7.11}$ & $62.95^{+6.62}_{-6.24}$ & $M_\odot$pc$^{-2}$\\ 
    \hline \hline
    \end{tabular}%
    }
          \\
          \justifying
          \small \vspace{9pt}
          \textbf{Notes.} The prior PDFs for both fiducial cases are the same: Gaussians centered at the parameters in the MS column with variance $\sigma_{prior} = 2$ with the corresponding units. The PDFs of the SFH are truncated at zero. The values for the inferred parameters are computed by fitting the \texttt{SciPy} Gaussian kernel-density estimator \citet{Pauli2020} to the accepted combinations of parameters in the last 70 ABC steps (see Sect. \ref{subsect:abc_parameters}). The posteriors shown in this table are presented considering the resulting most probable value of the distributions, as well as the 16$^{th}$ and 84$^{th}$ percentiles of each fitted parameter. 
\end{table*}

\begin{figure*}
    \centering
    \includegraphics[width=\textwidth]{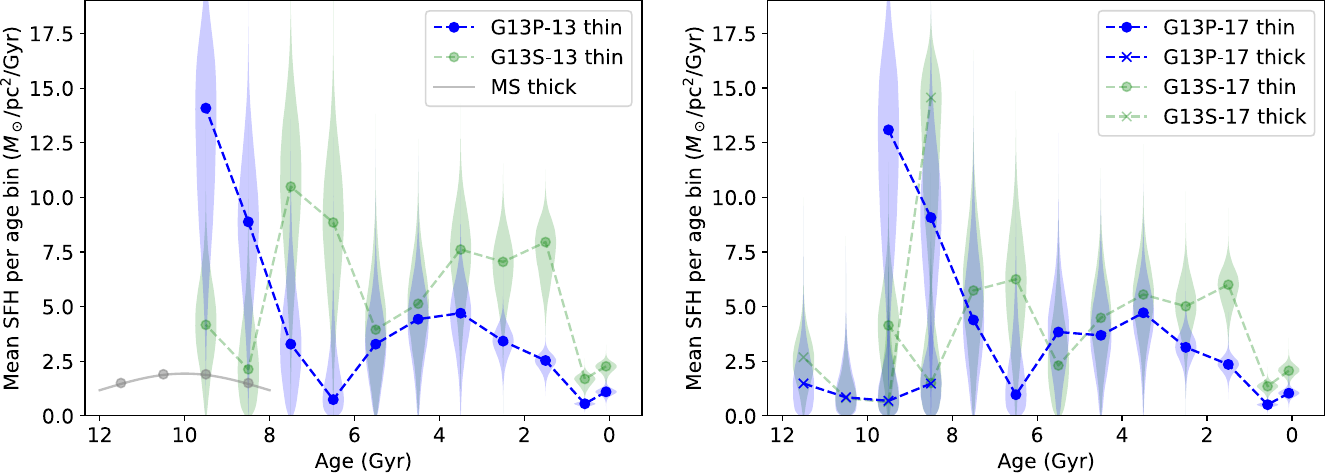}
    \caption{Posterior SFHs inferred with BGM FASt in the different executions. Those performed fixing the thick disc are found on the left, while the executions including the fit of the thick disc SFH are set on the right. For completeness, in the left plot, we also show in gray the fixed thick disc SFH. We distinguish in blue and green the stellar evolutionary model used to build the MS of each execution, PARSEC and STAREVOL, respectively. The marginalized posterior distributions for each of the weights of each age bin are represented as vertical shaded areas (the so-called violin plot).}
    \label{fig:SFH}
\end{figure*}

\begin{figure}
    \centering
    \includegraphics[width=\columnwidth]{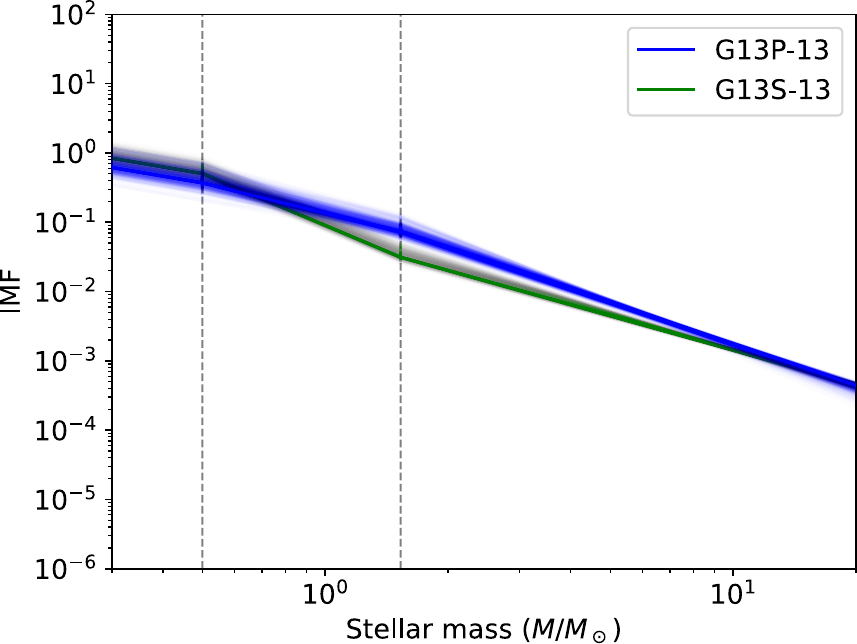}
    \caption{Posterior IMF inferred with BGM FASt in the executions G13P-13 (blue) and G13S-13 (green). In grey dashed lines are shown the limits of the three power-law IMF ranges, spanning $0.015-0.5 M_\odot$, $0.5-1.53 M_\odot$, and $1.53-120 M_\odot$. We recall that the value of the low-mass slope of the IMF is fixed to $\alpha_1 = 1.0$.}
    \label{fig:IMF}
\end{figure}

\subsection{Posterior PDFs of the inferred IMF and SFH parameters} \label{subsect:imf_and_sfh}

In Table \ref{tab:results} and Figs. \ref{fig:SFH} and \ref{fig:IMF}
we present the most probable values of the SFHs and IMF with their 16$^{th}$ and 84$^{th}$ percentiles. In Fig. \ref{fig:parameters_evolution} we show the evolution of the parameters for the execution G13P-13 over the 100 processed ABC steps (see Sect. \ref{subsect:abc_parameters}), which allows us to evaluate the degree of convergence (or divergence) of each of them. In Figs. \ref{fig:cornerplot_PARSEC} and \ref{fig:cornerplot_STAREVOL} we provide the corner plots with the approximate posterior PDFs of the two executions with the frozen thick disc SFH, which contain information on the correlations among the parameters. In Fig. \ref{fig:cornerplot_zoom} we zoom in on the most relevant features of these plots for those parameters directly related to the joint inference of the IMF and SFH processed here. Detailed comments on Figs. \ref{fig:parameters_evolution}, \ref{fig:cornerplot_PARSEC}, and \ref{fig:cornerplot_STAREVOL} can be found in Appendix \ref{app:evolution_parameters} and \ref{app:corner_plots}, respectively.  

\subsubsection{Initial mass function}

Fig. \ref{fig:IMF} shows that the IMF slopes $\alpha_2$ and $\alpha_3$ inferred in the G13P-13 execution lead to a smooth and continuous evolution across the entire mass range (blue lines and shaded areas). The inference of these parameters in G13S-13 (green lines), however, produces a non-gradual behavior around 1.5 $M_\odot$ ($\alpha_2>\alpha_3$). If real, this would imply a notable break in the IMF that has not been observed in earlier work and is hard to reconcile with current star-formation models.
We highlight that both inferences use the same input parameters in the MSs--except for the stellar evolutionary model, as well as the same set of priors. Furthermore, no significant correlations are found between $\alpha_2$ and $\alpha_3$ in either G13P-13 or G13S-13 (see Fig. \ref{fig:cornerplot_zoom}, top panel). More importantly, as shown in Fig. \ref{fig:parameters_evolution}, all the IMF parameters achieve a fast convergence during the ABC fit, always before the first 50 steps. Thus, the markedly different final most probable values obtained in G13P-13 and G13S-13 (see Table \ref{tab:results}) confirm the crucial role of the stellar evolutionary models in the inference process. Regarding the IMFs derived in executions G13P-17 and G13S-17, almost the same behaviours are observed with respect to G13P-13 and G13S-13, respectively. 

\subsubsection{The thin and thick discs' star formation histories} \label{subsect:results_imf_sfh}

In Fig. \ref{fig:SFH} we compare the SFHs resulting from the different BGM FASt executions: fixing the thick disc SFH on the left and fitting it on the right. All solutions have in common a tiny star formation history at early ages (<1 Gyr ago) followed by a wide enhancement in the star formation between 1-2 and 5-6 Gyr ago and a hiatus around $\approx 5-7$ Gyr ago. Each of these three main features of the thin disc evolution deserves special discussion. 
 
Although young stars (ages <1 Gyr) in our up to G=13 sample are distributed along the full range of masses, they are more concentrated at the upper main sequence, the region of the CMD where the inference of stellar masses and ages is most reliable. That explains why our strategy allows us to achieve very narrow distributions (small error bars) for the inferred parameters of the young thin disc SFH at ages between 0 and 2 Gyr ($\Sigma_\odot^0$, $\Sigma_\odot^1$, and $\Sigma_\odot^2$). We can state that all solutions confirm an abrupt decrease of the star formation in the Solar neighborhood approximately 1-1.5 Gyr ago.

As mentioned, all solutions confirm the presence of a wide enhancement of star formation at intermediate ages. However, while it spans from $\approx 2$ to $\approx 5$ Gyr ago in PARSEC solutions (G13P-13 and G13P-17), it is clearly shifted in about $\sim 1$ Gyr in STAREVOL executions (G13S-13 and G13S-17). Note that this shift supports the statement in \cite{Haywood2016} that the current uncertainty in the stellar evolutionary models at intermediate ages can trigger differences in the derivation of the SFH of, at least, $\sim 1-1.5$ Gyr. 

\begin{figure}[h!]
    \centering
    \includegraphics[width=0.7\columnwidth]{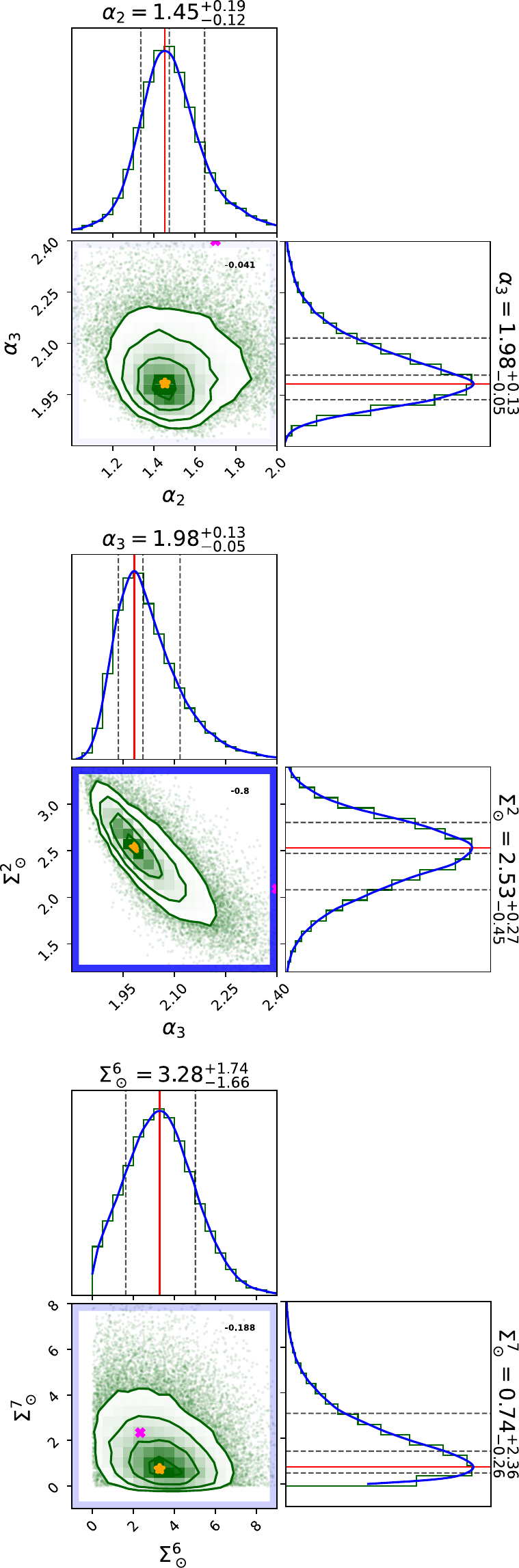}
    \caption{Correlations among some of the BGM FASt inferred parameters with their corresponding projected approximate posteriors PDF (with a Gaussian fit) from our fiducial execution G13P-13. The resulting most probable value and 16$^{th}$ and 84$^{th}$ percentiles are also marked with solid and dashed lines, respectively. It is also indicated with a dashed line the median of the distributions. In black, at the top right of each plot, we show Pearson’s correlation coefficient. High values of this coefficient are highlighted with an intense blue frame in the plot. Finally, the parameters adopted by the MS are marked with a magenta cross and the final values with an orange star. The full corner plot is presented in Appendix \ref{app:corner_plots}.}
    \label{fig:cornerplot_zoom}
\end{figure}

All solutions reproduce the hiatus in the SFH of the thin disc widely discussed in recent literature (see \citealt{Snaith2015}, subsequent studies and Sect. \ref{sect:sfh_imf_literature}). Furthermore, it appears more pronounced in the case of PARSEC solutions. We find this hiatus located at a lookback time of $\approx 6.5$ Gyr for PARSEC and $\approx 5.5 $ Gyr for STAREVOL. The correlation among points adjacent to the hiatus, which can be found in the mid panel of Fig. \ref{fig:cornerplot_zoom}, is very low. The interpretation of this hiatus in terms of Galactic disc evolution is discussed in Sect. \ref{sect:discussion}. 

At lookback times $> 6$ Gyr, PARSEC solutions present a pronounced enhancement of star formation. The STAREVOL run, although less pronounced, also detects this feature. However, at these older ages, this enhancement has to be discussed in the full scenario of the thin plus thick disc evolution. Our strategy, when considering only stars up to G=13, as done in this first paper, suffers from the degeneracy between the old stars in the CMD, especially the thin and the thick discs. In Fig. \ref{fig:SFH} we have added, to the thin disc SFH ($\Sigma^{0}_\odot$ to $\Sigma^{10}_\odot$, ages from 0-10 Gyr), the most probable values we derive for the four thick disc SFH parameters ($\Sigma^{9T}_\odot$ to $\Sigma^{12T}_\odot$, ages from 8-12 Gyr) in the right panel and the values used in the MSs in the left panel. As can be seen in Table \ref{tab:results}, the derived values for executions G13P-17 and G13S-17 remain very close to the initial priors, demonstrating that we cannot resolve the thick disc SFH at this limiting magnitude. This is justified by the lack of old stars at $G<13$ and the limitations of using CMDs as summary statistics without taking into account chemistry and kinematics (see Sect. \ref{sect:discussion}). Due to these limitations, in this work, we consider executions G13P-13 and G13S-13, obtained by fixing the thick disc SFH, more reliable than the ones derived fitting the CMD, and we will limit our analysis to them from now on in this article.

\subsubsection{Identification and impact of parameter correlations} \label{subsect:impact_correlations}

The consistency of these results requires a thorough and rigorous analysis of the corner plots (Figs. \ref{fig:cornerplot_PARSEC} and \ref{fig:cornerplot_STAREVOL}, and zoom images in Fig. \ref{fig:cornerplot_zoom}) and of the figures showing the evolution of convergence of the fitted parameters (Fig. \ref{fig:parameters_evolution}). The corner plots show different zones with non-negligible values of the Pearson coefficient ($R > 0.4$), which quantifies the linear correlation between two parameters. First, we must highlight both for G13P-13 and G13S-13 the important degeneracy between the third slope of the IMF and the recent SFH from 1 to 3 Gyr ago. See, for instance, $R(\alpha_3, \Sigma_\odot^2) = -0.800$ and $R(\alpha_3, \Sigma_\odot^2) = -0.898$ for G13P-13 and G13S-13, respectively. This is an important correlation given that the young ($\tau<3$ Gyr) and massive ($M>1.53 M_\odot$) stars are well-described at limiting magnitude $G=13$, representing 40\% of the sample in the MSs. In Sect. \ref{subsect:limits_and_caveats}, we provide insights on how to overcome this challenging degeneracy in upcoming works.

Secondly, we observe also in both solutions (even though more pronounced for G13S-13), correlations in the zone along the diagonal describing the posterior PDFs of consecutive--or second consecutive--parameters in the young SFH ($\tau<4$ Gyr). For example, we see correlations between $\Sigma_\odot^1-\Sigma_\odot^3$ or $\Sigma_\odot^2-\Sigma_\odot^4$, with $R = 0.404$ and $R = 0.441$ for G13P-13 and $R = 0.605$ and $R = 0.630$ for G13S-13. The smaller values of $R$ in this region for G13P-13 than for G13S-13 make us think that the PARSEC's solution for the young SFH is more robust. Although this type of correlation could be considered as something expected as consecutive parameters represent stars with similar features in the CMD, these results will require further analysis in future executions. 


We highlight a third region of the corner plots, which is the one described by $\alpha_3$ and the very old SFH ($\tau>9$ Gyr in G13P-13 and $6<\tau/\text{Gyr}<8$ in G13S-13). For G13S-13, this region of correlations can be extended for the whole combinations of $\Sigma_\odot^7$ and $\Sigma_\odot^8$ with the parameters describing the young SFH, mainly $\Sigma_\odot^1$ and $\Sigma_\odot^2$. If we compare the position of high-mass stars in the CMD with that of old stars, we find that a non-negligible number of both falls in the red clump (more than 900,000 stars with $M>1.53 M_\odot$ and around 700,000 stars with $\tau>8$ Gyr in G13P-13). Young--and massive--stars and old stars clearly constitute two different populations in the Galaxy. The observed correlations in this region demands for including additional sources of information in the future, such as chemistry or kinematics, in order to break the current degeneracies in the CMD.

\begin{figure*}[ht]
    \centering
    \includegraphics[width=\textwidth]{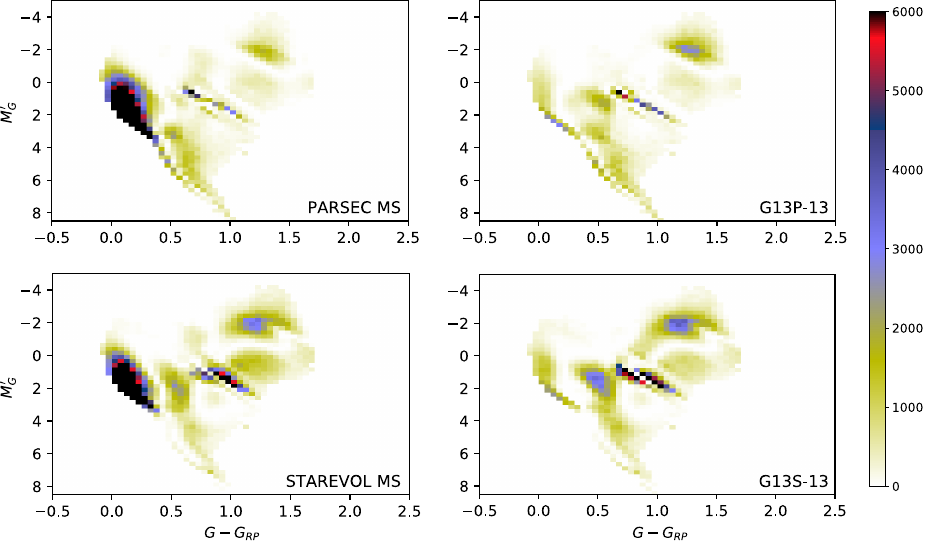}
    \caption{Low-latitude ($|b|<10$) color-magnitude diagrams illustrating the distance per bin, as defined in Eq. (\ref{eq:distance}), for both the initial MSs (left) and the final solutions (right) of the BGM FASt executions G13P-13 (top) and G13S-13 (bottom).}
    \label{fig:CMD_zoom}
\end{figure*}

\begin{figure}[ht]
    \centering
    \includegraphics[width=\columnwidth]{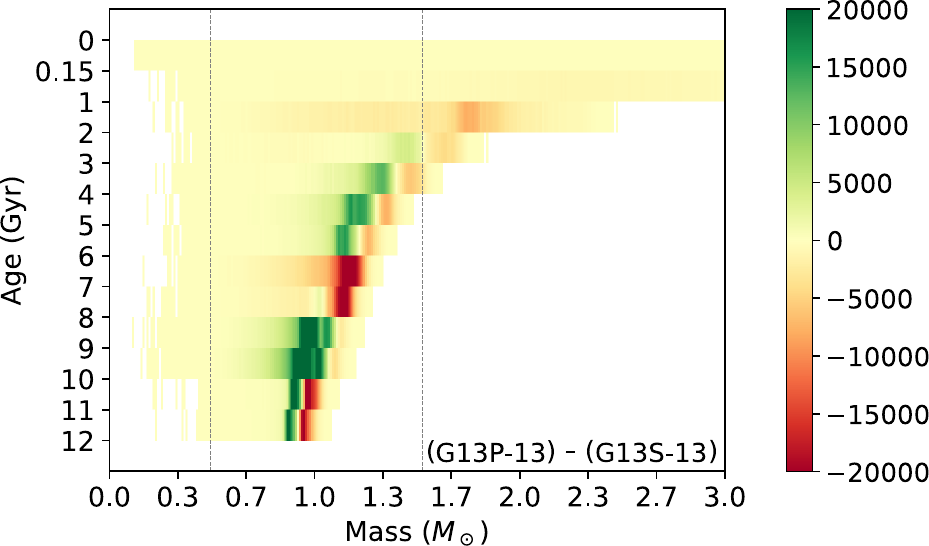}
    \caption{Bin-wise differences in stellar number density in mass-age space between the two final solutions obtained considering a fixed thick disc SFH, that is, G13P-13 and G13S-13. For reference, we show in the mass axis the limits of the three power-law IMF ranges shown in Table \ref{tab:results} (grey dashed vertical lines).}
    \label{fig:final_mass_age_space}
\end{figure}

\subsection{BGM CMDs versus {\it Gaia} data: an evaluation of the stellar evolutionary models} \label{subsect:cmds_evolution}

In Appendix \ref{app:complete_cmds} (Figs. \ref{fig:CMD_PARSEC} and \ref{fig:CMD_STAREVOL}) we present the complete set of CMDs describing the comparison between the MSs and the final BGM FASt PSs (top and bottom panels of each figure, respectively) with the {\it Gaia} data for executions G13P-13 and G13S-13. To provide a more focused analysis, Fig. \ref{fig:CMD_zoom} highlights the most relevant subplots of the entire set of CMDs. These plots allow us to analyze the regions of the CMD in which the new IMF and SFH significantly improve the resemblance with the {\it Gaia} catalog. This process also allows us to assess the caveats of this method and the still most significant discrepancies between {\it Gaia} and the model.  

The evolution of the distance metric defined in Eq. (\ref{eq:distance}) over the inference process (similarity between catalog--initial MS and catalog--final PS) is presented in Table \ref{tab:numbers_and_distances}. For G13P-13, we start with an initial value of 1,772,695 and it is reduced to a residual distance of 922,755, which implies an improvement of almost factor 2. The outcome of BGM FASt G13S-13 yields a less favorable solution. It begins with a higher initial distance to {\it Gaia} for the MS (1,825,293) that is reduced by only a factor 1.3 after applying BGM FASt (1,362,537), far from the clear improvement in G13P-13.
Regarding the analysis of the CMDs, despite the overall agreement, several regions show a discrepancy between the model and the data. The discrepancies can be seen in the third column of Figs. \ref{fig:CMD_PARSEC} and \ref{fig:CMD_STAREVOL}, where we present the distance for each given bin (higher the distance, larger the discrepancy), and/or in the fourth column, where the color represents the simple difference between the model and the data star counts. Focusing first on the CMDs of execution G13P-13 (Fig. \ref{fig:CMD_PARSEC} with a zoom in Fig. \ref{fig:CMD_zoom}), it is interesting to compare the discrepancies between the initial MS and {\it Gaia} (top) with the final discrepancies after the BGM FASt execution (bottom):

\begin{itemize}
    \item Upper main sequence: this region initially showed a significant excess of stars in the MS, especially visible at low and intermediate latitudes (as expected for massive stars). After the fit, the discrepancy is almost completely resolved thanks to the new IMF and SFH, responsible for reducing the number of stars from  8,572,604 in the MS to 7,113,842 in the final PS, much closer to the value of 7,266,095 stars present in the {\it Gaia} $G<13$ sample. However, a slight shift of the entire main sequence in color is still present. This region is where BGM FASt has been most efficient.
    \item Asymptotic giant branch (AGB): in this region, the distance $\delta_p$ shows large values, mostly at intermediate and low latitudes, and does not improve much after the fit. Comparing the first and second columns of Fig. \ref{fig:CMD_PARSEC}, we see that the upper AGB is absent in the MS. This is due to the tracks, that do not cover well this region, so the fit cannot improve the resemblance between the simulations and the catalog in it. Furthermore, in the fourth column, we observe that the absolute differences in star counts are low compared to the significant number of stars present in the main sequence and the red giant branch. Therefore, this region has no impact in the fit.
    \item Red clump: the MS produces a red clump that is roughly positioned correctly but appears less compact (more extended in color) than in the catalog. The new IMF and SFH fit improves this region, particularly at high and mid latitudes. However, the discrepancies persist at lower latitudes, which could be partially due to some remaining inaccuracies in the extinction map used in the MS or/and to nearby star forming regions.
    \item Subgiants and lower main sequence: these regions initially exhibited a slight lack of stars in the MS, which has been reduced after the fit. Although the lack is still present, we suspect it could be completely removed if the color shift is corrected.
\end{itemize}

In the case of using STAREVOL (Fig. \ref{fig:CMD_STAREVOL}), similar patterns are seen, but the discrepancies are generally stronger. Notably, one of the major issues with these tracks is the position of the red giant branch, which is redder in the simulation than in the data. This discrepancy cannot be corrected with our process, likely contributing to the lower accuracy of the fit obtained with STAREVOL. 

Complementing the physical information presented in the CMDs, the differences in the mass-age distribution resulting from the G13P-13 and G13S-13 executions are shown in Fig. \ref{fig:final_mass_age_space}. As can be seen, the discrepancies between the two stellar evolutionary models show a non-continuous trend. We have to highlight that this behavior is reflected in all processes of IMF and SFH parameter inference and it is the consequence of two effects: 1) the differences in the tracks of the PARSEC and STAREVOL evolutionary models and 2) the lack of correspondence between the IMF and SFH derived in each case (see Table \ref{tab:results} and Fig. \ref{fig:SFH}). While the effect of the particular tracks is difficult to understand in the mass-age distribution, it is relatively easy to observe the consequences of using different IMFs or SFHs in this space. For example, we clearly see the shift of 1-1.5 Gyr commented in Sect. \ref{subsect:results_imf_sfh}.

\section{Discussion} \label{sect:discussion}

In Sect. \ref{sect:sfh_imf_literature} we compare our findings on the IMF and SFH  to recent studies in the literature. Sect. \ref{subsect:limits_and_caveats} presents a deep analysis of the caveats and limitations of the present work and advances new strategies under development. 

\subsection{IMF and SFH of the Solar vicinity} \label{sect:sfh_imf_literature}

The initial mass functions (IMF) and its critical role in many fields of stellar, Galactic, and extragalactic astrophysics are still a matter of debate within the community. The results presented in this work (see Sect. \ref{subsect:results_imf_sfh}) illustrates how the derivation of this fundamental function critically depends on the accuracy of the stellar evolution models being used. Our IMF derivation, based on a full-sky sample up to magnitude $G=13$ where $\sim 60$\% of the stars fall in the range of masses 0.5-1.53 $M_\odot$, suggests that, apart of the dependency on the stellar evolutionary models, the values for $\alpha_2$ derived here are statistically robust when compared with other recent estimates. Furthermore, our derivations for $\alpha_3$ have a high added value thanks to  the high number of massive stars in our limiting magnitude sample compared to other solutions (these stars are intrinsically bright so we are considering here all the massive {\it Gaia} sources up to approximately 1-2 kpc from the Sun). Nonetheless, as discussed in Sect. \ref{subsect:impact_correlations}, the high correlation between $\alpha_3$ and $\Sigma_\odot^2$ demands further developments (see Sect. \ref{subsect:limits_and_caveats}). 

\begin{figure*}[ht]
    \centering
\includegraphics[width=\textwidth]{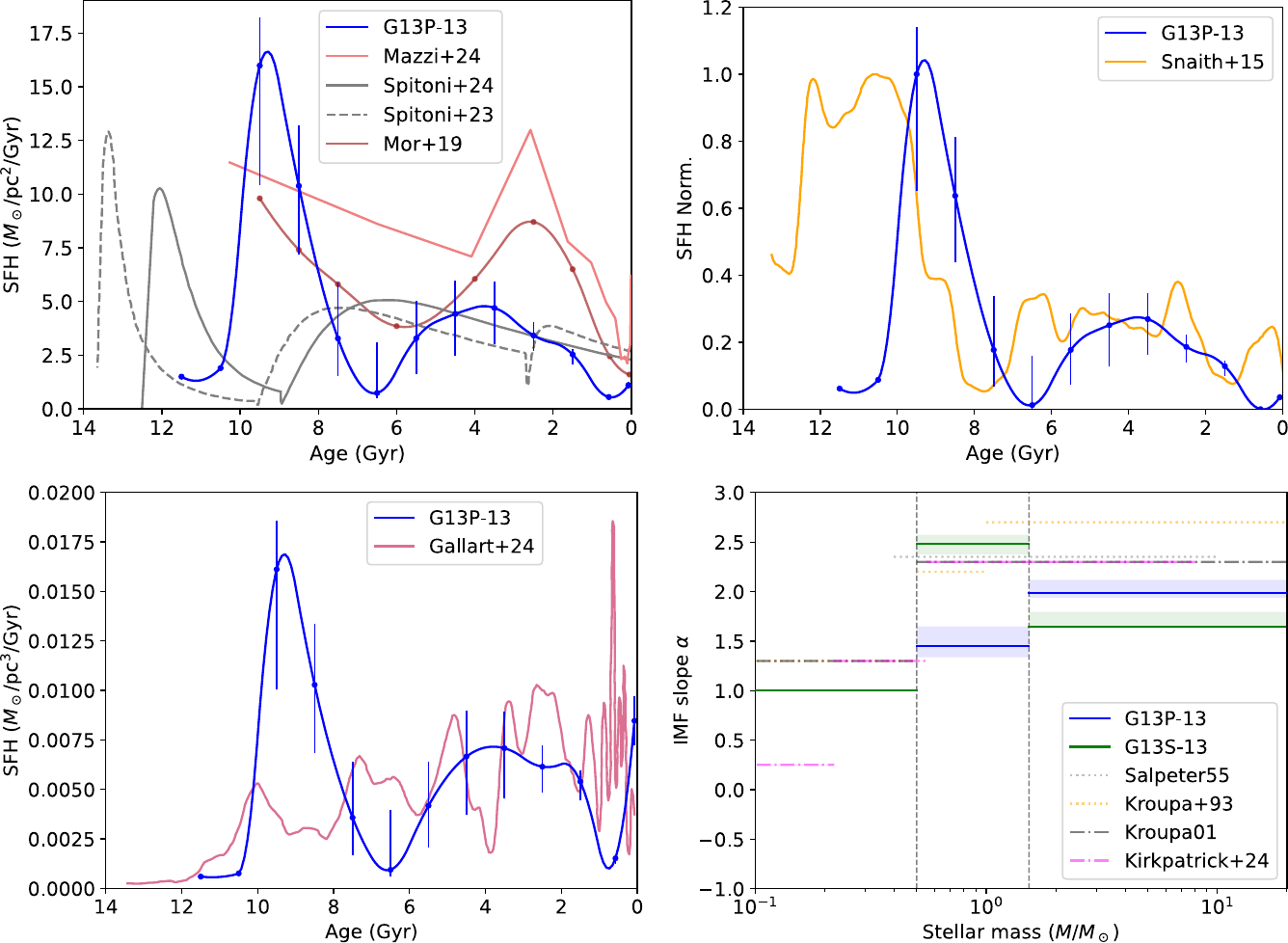}
    \caption{Comparison of our inferred SFH and IMFs with the literature. \textbf{Top left}: we compare our fiducial solution G13P-13 with the results from \cite{Mazzi2024}, \cite{Spitoni2024}, \cite{Spitoni2023}, and \cite{Mor2019}, all presented in original units (surface density). The error bars of our SFH and IMFs correspond to the 16$^{th}$ and 84$^{th}$ percentiles of the marginalized posterior PDFs of each parameter, whose values can be found in Tab. \ref{tab:results}. \textbf{Top right}: the findings of \cite{Snaith2015} are contrasted with our fiducial solution normalized to the maximum value of each SFH. \textbf{Bottom left}: we compare the solution of our execution G13P-13 with the SFH from \cite{Gallart2024}, presented in terms of volume density. To convert our surface densities to volume densities we employ the scale heights of each population in BGM. \textbf{Bottom right}: the slopes of the IMFs derived in executions G13P-13 and G13S-13 are shown alongside those from \cite{Salpeter1955}, \cite{Kroupa1993} (assumed in \citealt{Gallart2024}), \cite{Kroupa2001}, and \cite{Kirkpatrick2024}. Dashed vertical lines delineate the mass ranges of this work, as detailed in Table \ref{tab:results}.}
        \label{fig:sfh_imf_lit}

\end{figure*}

In the bottom right panel of Fig. \ref{fig:sfh_imf_lit} we compare our G13P-13 and G13S-13 IMFs with widely used values from the literature and recent estimates. We find that, in all the works shown in the figure except for the STAREVOL solution G13S-13, the slopes of the IMF progressively grow with mass ($\alpha_1<\alpha_2<\alpha_3$). As mentioned in Sect. \ref{subsect:imf_and_sfh}, the fact that $\alpha_3<\alpha_2$ in G13S-13 could be one of the drawbacks of this execution. The value of the second IMF slope we obtain in G13P-13 for the mass range 0.5-1.53 $M_\odot$ has small error bars (see Table \ref{tab:results}) and it is not correlated with other parameters (see Fig. \ref{fig:cornerplot_PARSEC}). This value is slightly below the commonly used slopes from \cite{Kroupa2001} and \cite{Salpeter1955} (see Fig. \ref{fig:sfh_imf_lit}). For masses larger than 1.53$M_\odot$, the value of the IMF slope derived using PARSEC (G13P-13) is consistent with the Salpeter's IMF and with recent  values derived from completely independent methods like, for example, that obtained by \cite{Dickson2023} fitting a multimass model to globular clusters data.

Let us now analyse results for the star formation history of the Solar vicinity, whose derivation has seen a tremendous revival in recent times following the publication of large astrometric, spectroscopic, and photometric surveys. Intending to provide a synthetic view of the local SFH, in Fig. \ref{fig:sfh_imf_lit} we compare the results obtained using different methods.
We believe that the representative compilation shown in these figures present many of the features that are currently under debate.
The figures include the results of three different approaches to the problem. One of the techniques pursue fitting Galactic chemical evolution models to the abundances derived from spectroscopic data (e.g. \citealt{Snaith2015, Haywood2016, Spitoni2023, Palla2024}). Another consists on mimicking the color-magnitude diagrams obtained from photometric observations (in recent years, mainly {\it Gaia} data) by considering population synthesis models of the Galaxy (e.g. \citealt{Mor2019, Mazzi2024, Gallart2024}, and the present approach). All studies based on {\it Gaia} CMDs provide estimates of a dynamically evolved SFH (that is, an SFH convolved with the effects of stellar dynamical mixing) rather than the true SFH, as explained in e.g. \citet{Gallart2024}. A third group aims to derive the SFH using specific tracers, like the white dwarf population. As detailed in the recent review by \cite{Tremblay2024}, local volumes of white dwarfs are essential for studies of Galactic star formation history. 


In Fig. \ref{fig:sfh_imf_lit} we provide our fiducial PARSEC SFH solution together with literature values in three different ways: the top left panel shows the comparison with works that derive the SFH in units of surface density (M$_{\odot}\, {\rm pc}^{-2}\, {\rm Gyr}^{-1}$); the top right panel shows comparisons with normalized (relative) SFHs; finally, the bottom left panel shows comparisons in terms of the SFH per local volume density (in M$_{\odot}\, {\rm pc}^{-3}\, {\rm Gyr}^{-1}$). 

Overall, we observe how most of the recent derivations of the local SFH present a first enhancement of star formation at intermediate ages followed by a pronounced hiatus that is succeeded by a complex and discrepant evolution at ages older than 8 Gyr, just when the thick disc begins to play a relevant role. More in detail, as can be seen in Fig. \ref{fig:sfh_imf_lit}, both the width of the enhancement and the location of the hiatus change from one solution to another. In the following paragraphs, we attempt to discuss how different models, assumptions, sets of data, and/or fitting strategies contribute to the complex understanding of these discrepancies. 

\cite{Snaith2015} derive an SFH for the inner disc ($R<10$ kpc) by fitting a closed-box-like system in this region to the age-[Si/Fe] relation of the HARPS GTO high-resolution ($R=110,000$) spectra sample of FGK stars from \cite{Adibekyan2012}, with ages determined in \cite{Haywood2013}. According to their model, most accretion takes place at early times, when substantial star formation has not occurred yet. Therefore, only a single infall of gas is considered in this case in the very early times of the Galaxy, and the later SFH variations are a consequence of its internal evolution, e.g. the formation of the bar and/or the end of the thick-disc phase \citep{Haywood2016}. According to the authors, the cessation of star formation during a period of $\sim 1-2$ Gyr could reproduce the bimodal distribution found in the [Fe/H]-[$\alpha$/Fe] diagram of APOGEE data, among other spectroscopic surveys \citep{Haywood2016}. Looking at Fig. \ref{fig:sfh_imf_lit} (top right) our fiducial PARSEC SFH matches their normalized SFH in the interval 1-10 Gyr ago with only a shift of 1-2 Gyr. 

Different approaches are taken in \cite{Spitoni2023} and \cite{Spitoni2024}. Both works consider Galactic chemical models in which different infalls of metal-poor gas from external accretion could be responsible for the distinguished high- and low-alpha sequences. \cite{Spitoni2023} compare their three-infall model to the {\it Gaia} DR3 GSP-Spec chemical abundances for $\alpha$ elements, focusing their efforts on the characterization of the population of young and massive stars with a deficiency in metallicity [M/H] in the Solar neighborhood (guiding radius between 8.1 and 8.4 kpc). In contrast, \cite{Spitoni2024} adjusts a two-infall chemical model to the [Fe/$\alpha$] vs. [$\alpha$/H] relation of a sample of red giant stars in the Solar neighbourhood (galactocentric distances between 7 and 9 kpc and $|z|<2$ kpc) observed by APOGEE. It is worth mentioning that none of the previous works include stellar migration, a process that is instead considered in the three-infall model of  \cite{Palla2024}, leading to similar results as the three-infall model of \cite{Spitoni2023}. An interesting feature of \cite{Palla2024} is that they use open clusters to study the variation of the SFR with the Galactic radius. All in all, when comparing our G13P-13 SFH to the solutions arising from the chemical evolution models of \cite{Spitoni2023}, \cite{Spitoni2024} and \cite{Palla2024}, and also with that of \cite{Snaith2015} discussed above, we can confirm that our approach, different both in modeling and data, stands coherent with the existence of the hiatus, also unveiled in our previous fit to the {\it Gaia} DR2 up to $G=12$ catalog \citep{Mor2019}.



The comparison of our results with those from \cite{Mazzi2024} and \cite{Gallart2024}, both based on CMD-fitting techniques, is difficult due to the lack of detail of the former at ages older than 2-3 Gyr (they fit a logarithmic SFH instead of a linear one) and the high level of discretization at recent times for the latter. It is especially interesting to contrast our inference with that of \cite{Gallart2024}. Although based on a similar methodology--a CMD-fitting process applied to {\it Gaia} data, some of the strategies and basic ingredients used in their work are significantly different than ours. Let us mention some of them and, more importantly, discuss the scientific outputs these differences may entail: 1) \cite{Gallart2024} used a volume complete sample restricted to the 331,312 stars currently in the 100 pc Solar neighborhood (GCNS), whereas our all-sky catalog, also based on a well-defined observational cut (apparent limiting magnitude $G=13$), includes a factor 20 more stars ($\sim$7 million sources); 2) while \cite{Gallart2024} impose the \cite{Kroupa1993} initial mass function, the simultaneous fit of the IMF and the SFH done in this work may be crucial to successfully understand the dependencies and degeneracies among both fundamental quantities. For instance, as seen in the bottom right panel of Fig. \ref{fig:sfh_imf_lit}, the high value of IMF slope at masses $>1 M_\odot$ used by \cite{Gallart2024} could be behind their large values of the SFH at recent times ($\tau<2$ Gyr); and 3) as deeply discussed in Sect. \ref{subsect:cmd_description}, we know that systematic differences are also present due to the different stellar evolutionary models used, PARSEC and STAREVOL in our case and BASTI-IAC in \cite{Gallart2024}. 
Even considering all the effects mentioned so far, further work is needed to justify the significant differences observed in the bottom left panel of Fig. \ref{fig:sfh_imf_lit}. 

Our fitting technique demonstrates to be statistically robust when deriving the more recent phases of disc evolution. In particular, we observe a sudden decrease of star formation in the last 3 Gyr, and the SFH values describing this behavior present tiny error bars (see Fig. \ref{fig:SFH}). This trend seems to be, in terms of shape, in agreement with the star formation history determined by \cite{Mazzi2024}, which is the result of collapsing the different SFHs as a function of the distance from the Galactic plane obtained from fitting the {\it Gaia} DR3 CMD within a cylinder of 200 pc radius centered on the Sun and spanning 1.3 kpc above and below the plane. However, the total surface mass density obtained in their work, $118.7\pm 6.2 M_\odot \mathrm{pc}^{-2}$, appears much larger than ours ($\Sigma_\odot^T = 53.71^{+5.48}_{-6.89} M_\odot \mathrm{pc}^{-2}$ for G13P-13) and other works in the literature (see the area under the curves in top-left panel of Fig. \ref{fig:sfh_imf_lit}). There are several aspects to take into account to understand these significant discrepancies. In our view, the derivation of this parameter requires the full achievement of the dynamical self-consistency in the process, a challenge that we plan to address through the iteration between the BGM FASt solutions for the IMF and SFH, and the fit to the gravitational potential to kinematic and densities from Gaia data (see Fig. 2 from \citealt{Robin2022}). Until this is achieved, estimates of this parameter can suffer for significant biases coming from the combined use of different volume samples and simplified expressions for the radial or vertical spatial distributions. As an example, as found in \cite{Robin2022}, a fitting to kinematic Gaia DR2 data indicates that the effective radial scale length of a population varies with the distance from the Galactic plane. Facing this would require a more complex modeling than classical exponential or sech2 vertical scale heights, which could affect the vertical structure of the SFH infered by \cite{Mazzi2024}. Additionally, recent tools derived by the GaiaUnlimited team (e.g. \citealt{Castro-Ginard2023}) shall be considered to correct the survey selection function when using different volume samples.

Regarding the present-day SFH in the Solar neighborhood, the Galactic chemical evolution model proposed by \cite{Prantzos2018} predicts a value of 2-5 $M_\odot \mathrm{pc}^{-2} Gyr^{-1}$ (see mid panel of Fig. 9 in the authors' article). Independently of the stellar evolutionary model being used, our results indicate an SFR of about 1.3-2.5 $M_\odot \mathrm{pc}^{-2} Gyr^{-1}$ for the last Gyr (see in Table \ref{tab:results} values for $\Sigma_\odot^0$ and $\Sigma_\odot^1$), very close to the value derived in \cite{Mor2019}. Furthermore, our results show a fast and stable convergence of these parameters in all BGM FASt executions (see Fig. \ref{fig:parameters_evolution}). Recent estimations from \cite{Mazzi2024} and \cite{Spitoni2023, Spitoni2024} point in the same direction, deriving a present-day of the SFH close to the lower limit of the wide range of $2-5 M_\odot \mathrm{pc}^{-2} Gyr^{-1}$ from \cite{Prantzos2018}.

\subsection{Limits of our analysis} \label{subsect:limits_and_caveats}

In this section, we summarise the main caveats of our analysis and how can we overcome them in the future. First and foremost, as discussed in Sect. \ref{subsect:cmd_description}, the choice of the stellar evolutionary models introduces a considerable systematic uncertainty, even when rather canonical models are used. This work has been a first approximation to the problem, but we are aware that a more thorough analysis must be done considering alternative tracks such as BASTI-IAC and PARSEC v2.0, which we will implement in future executions of BGM FASt. 

Second, as shown in Sect. \ref{subsect:results_imf_sfh}, some parameters like the SFH of the old thin disc are highly degenerated with the parameters of the thick disc SFH, and these correlations are impossible to break with the magnitude-limited sample $G<13$ used here. Furthermore, parameters related to old stars in general ($\tau>8$ Gyr) suffer from a lack of convergence throughout the ABC process (see Fig. \ref{fig:parameters_evolution}) and, as demonstrated in Fig. \ref{fig:cornerplot_zoom} and Sect. \ref{subsect:impact_correlations}, some of them are clearly correlated. The effects of degeneracy, correlation, and convergence are usually neglected in similar works,  resulting in underestimated uncertainties for the derived SFH. In conclusion, the current executions of BGM FASt up to $G=13$ does not allow us to break the degeneracies and correlations found in the earliest phases of disc evolution (stars with ages $> 8$ Gyr). We will tackle this problem in the future by considering samples up to further limiting magnitudes (including more old stars) and additional information such as chemistry or kinematics (better characterizing the populations). This approximation will also be useful to tackle the current degeneracies between $\alpha_3$ and the young stars SFH (see Sect. \ref{subsect:impact_correlations}).

Third, as mentioned in Sect. \ref{subsect:parameter_space}, throughout the BGM FASt inference we break the dynamical self-consistency of the MSs, which is clearly seen in Table \ref{tab:results} comparing the total surface densities of the MSs and the final solutions. We are working in two different possibilities to overcome this problem: 1) the iteration between BGM Std and BGM FASt until reaching a dynamical self-consistent solution with the best IMF and SFH; 2) the inclusion of new constraints in BGM FASt to maintain the dynamical self-consistency in the process.

\section{Conclusions and future work} \label{sect:conclusions}

We present the joint inference of the star formation history (SFH) and the initial mass function (IMF) of the Galactic disc in the Solar neighborhood obtained by iteratively fitting our dynamically consistent Besançon Galaxy Model (BGM) to the {\it Gaia} DR3 color-magnitude diagram (CMD) up to $G=13$. The process is carried out using an improved version of the BGM Fast Approximate Simulations (BGM FASt) tool developed within our team. Given the nature of the process, the inference of these two Galactic properties--the SFH and the IMF--has been done by deeply analyzing the critical role of the adopted stellar evolutionary models (SEMs) in our population synthesis model. We consider the derivation with PARSEC SEM to be our fiducial solution due to its significantly higher likelihood compared to the STAREVOL solution.

From {\it Gaia} DR3 and BGM FASt up to $G = 13$ we report a new derivation of the slopes of the IMF in the range [0.5, 120] $M_\odot$. Although statistically robust, they appear discrepant using different SEMs. In our fiducial execution with PARSEC, for the range 0.5-1.53 $M_\odot$ the slope takes a value of $\alpha_2 = 1.45^{+0.19}_{-0.12}$, while for masses larger than 1.53 $M_\odot$ we obtain $\alpha_3 = 1.98^{+0.13}_{-0.05}$. Using STAREVOL, the inferred values are $\alpha_2 = 2.48^{+0.09}_{-0.11}$ and $\alpha_3 = 1.64^{+0.15}_{-0.02}$. These parameters are derived with a remarkable level of convergence and high statistics due to the nature of the $G<13$ sample. For the SFH of the Galactic disc at the Solar neighbourhood, all our solutions have in common a clear hiatus around $\approx$ 5-7 Gyr ago with $\approx$1 Gyr shift depending on the evolutionary model. The presence of a secular mechanism is supported by the wide plateau in the bump between $\approx$6 and $\approx$2 Gyr ago, and the high convergence achieved for the young SFH confirms an abrupt decrease of the star formation in the Solar neighborhood approximately 1-1.5 Gyr ago. The fact that the statistically robust SFH derived here has several features in common with recent determinations in the literature (the hiatus, the burst, and the abrupt recent decrease) indicates our SFH is ready for the complex astrophysical interpretation of the disc evolution. 

In this work, we have identified the areas in the color-magnitude diagram where PARSEC and STAREVOL match {\it Gaia} or need improvement. In all BGM FASt executions, we tackle the initial discrepancies at the upper main sequence and the subgiants region, while issues in the asymptotic giant branch and the red clump seem to be related to the stellar tracks and therefore not solvable when fitting the IMF and the SFH. PARSEC stellar evolutionary models produce smaller final distance metrics with respect to {\it Gaia} whereas non-continuous features are observed as residuals of the fit in the case of STAREVOL.
   
The future BGM FASt executions being prepared pursue the goal of achieving a robust fitting of the SFH at old ages. For that, the current implementation of the code is ready to significantly increase the limiting apparent magnitude of the sample and, similarly to other analyses, to take into account the selection function of the {\it Gaia} sample \citep{Cantat-Gaudin2023, Castro-Ginard2023}. The code is also ready to evaluate and even derive the stellar binary fraction, another key ingredient that plays a crucial role in the dynamical evolution of any stellar system. In parallel, our new BGM FASt executions will require reintroducing the  BGM dynamical self-consistency. This will be tackled by incorporating spatial density and kinematics constraints. Last but not least, future BGM FASt executions must incorporate the use of chemical and kinematical data to address the complex evolution of the old thin and thick discs. All this work will be developed in parallel with the use of BGM FASt as a tool for testing, not only new SFHs and IMFs coming from literature but also other key ingredients of the population synthesis models arising in the near future as by-products of new {\it Gaia} Data Releases.


\begin{acknowledgements} We thank the anonymous referee for a constructive report that helped to revise and improve the quality of the manuscript. This work has made use of data from the European Space Agency (ESA) mission {\it Gaia} (\url{http://www.cosmos.esa.int/gaia}), processed by the {\it Gaia} Data Processing and Analysis Consortium (DPAC, \url{http://www.cosmos.esa.int/web/gaia/dpac/consortium}). Funding for the DPAC has been provided by national institutions, in particular the institutions participating in the {\it Gaia} Multilateral Agreement. This work was (partially) supported by the Spanish MICIN/AEI/10.13039/501100011033, by "ERDF A way of making Europe" by the “European Union” through grant PID2021-122842OB-C21, by the Institute of Cosmos Sciences University of Barcelona (ICCUB, Unidad de Excelencia ’Mar\'{\i}a de Maeztu’) through grant CEX2019-000918-M, by the OCRE awarded project Galactic Research in Cloud Services (Galactic RainCloudS, funding from the European Union’s Horizon 2020 research and innovation programme under grant agreement no. 824079) and by MCIN with funding from European Union NextGenerationEU(PRTR-C17.I1) and by Generalitat de Catalunya. All BGM simulations and part of BGM FASt fits were executed on computers from the Utinam Institute of the Université de Franche-Comté, supported by the Région de Franche-
Comté and the Institut des Sciences de l’Univers (INSU). FA acknowledges financial support from MCIN/AEI/10.13039/501100011033 and European Union NextGenerationEU/PRTR through grant RYC2021-031638-I.
\end{acknowledgements}

%
%
\bibliographystyle{aa}
\bibliography{references}

\begin{appendix} 

    \section{Main characteristics of the PARSEC and STAREVOL stellar models} \label{app:stellar_models}
    
    Comparing PARSEC and STAREVOL evolutionary models, one notices small differences in the treatment of the convection, both using the Mixing Length Theory, but with different $\alpha_{MLT}$ of 1.74 in PARSEC and 1.6264 in STAREVOL. Overshooting is taken into account in PARSEC from the convective core, not in STAREVOL. None of the two sets of stellar models implemented in BGM includes the rotation. The mass loss coefficient $\eta$ \citep{Reimers1975} is 0.5 in STAREVOL and 0.2 in PARSEC. About the Solar abundances, they differ slightly, being from \cite{Caffau2011} in the case of PARSEC and from \cite{Asplund2009} in STAREVOL. For the initial helium abundances, the values are slightly different as well, PARSEC using the formula $Y = 0.2485+1.78 \times  Z$, while STAREVOL assumes $\Delta Y/\Delta Z = 1.29$.
    
    \section{Evolution and convergence of the BGM FASt fitted parameters} \label{app:evolution_parameters}

\begin{figure*}[h!]
        \centering
        \includegraphics[height=0.95\textheight]{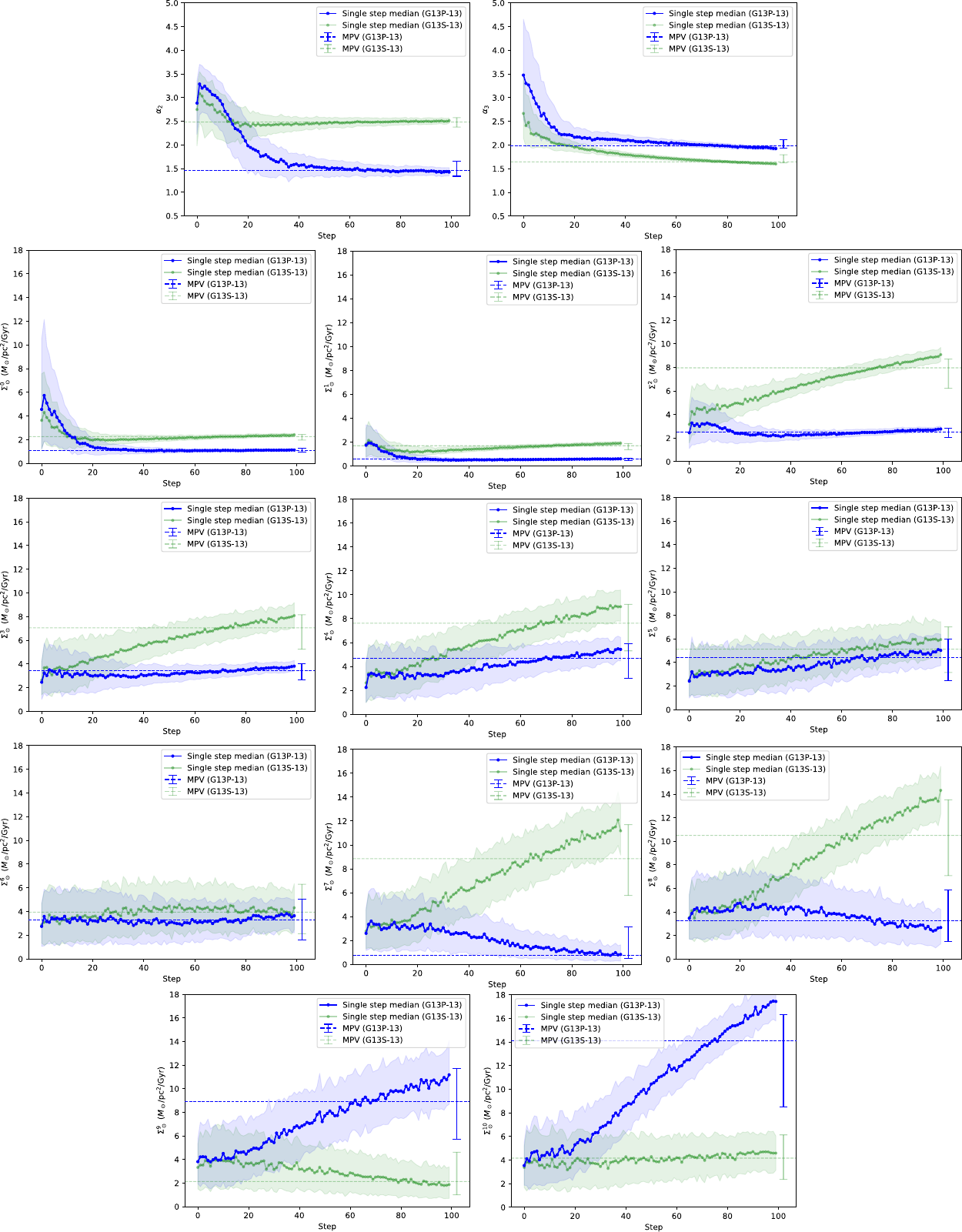}
        \caption{Evolution of the median at each step of the 13 BGM FASt inferred parameters across the 100 ABC steps (refer to the derivation process in Sect. \ref{subsect:cmd_technique}) for executions G13P-13 (blue) and G13S-13 (green). At the end of each plot, we present the final most probable values (MPVs) of these parameters, along with their 16$^{th}$ and 84$^{th}$ percentiles as determined in Sect. \ref{subsect:abc_parameters}.}
        \label{fig:parameters_evolution}
    \end{figure*}

    We show in Fig. \ref{fig:parameters_evolution} the evolution of the 13 parameters inferred in executions G13P-13 (blue) and G13S-13 (green) over the 100 ABC steps, as well as their final most probable values (MPV) with 16$^{th}$ and 84$^{th}$ percentiles as error bars. In the case of G13P-13 solution (mother simulation built using PARSEC stellar tracks and thick disc SFH not considered in the inference), we observe a strong convergence of the two slopes of the IMF ($\alpha_1$ and $\alpha_2$), as well as the parameters describing the young SFH up to 3 Gyr ago. We don't find strong trends for $\Sigma_\odot^4$, $\Sigma_\odot^5$, even though we cannot consider these parameters effectively converged. In the case of $\Sigma_\odot^6$, the stability of its MPV contrasts with its non-negligible error bars. The most important problems appear for the SFH parameters describing stars with $\tau>6$ Gyr, for which we find a lack of convergence and important trends. As explained in Sect. \ref{subsect:limits_and_caveats}, we propose this problem to be related with the lack of information we have for old stars at $G<13$ and using only photometric data.

    Regarding the convergence of the parameters for G13S-13 (same as G13P-13 but using STAREVOL stellar tracks to build the mother simulation), we find a similar picture as in G13P-13 for the two slopes of the IMF, the very young SFH ($\tau<1$ Gyr) and the parameters describing the old SFH ($6>\tau/\text{Gyr}>8$ Gyr). However, in this case we observe a worrying steep evolution of SFH parameters for stars with $1<\tau/\text{Gyr}<4$. On the other hand, more stable values are found for $\Sigma_\odot^9$ and $\Sigma_\odot^{10}$, even though their important uncertainties are an evidence of a lack of characterization of stars at those ages.
    
    \section{Corner plots of the IMF and SFH inferred parameters} \label{app:corner_plots}
    
\begin{figure*}[h!]
        \centering
        \includegraphics[width=\textwidth]{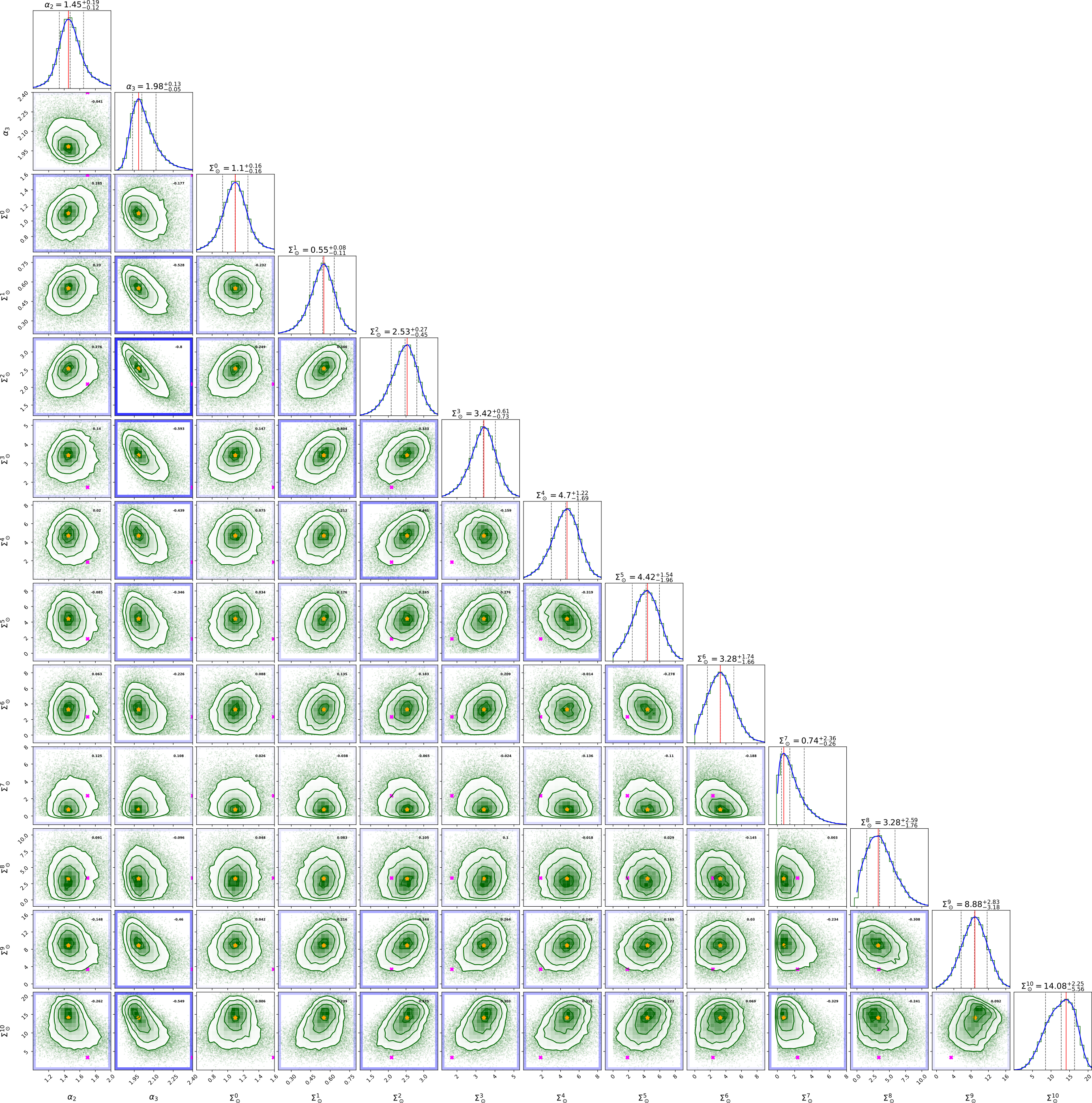}
        \caption{Corner plot containing the parameters derived from our fiducial execution G13P-13, including two slopes of the IMF and 11 parameters for the SFH of the thin disc. At the top of the columns, the projected approximate posterior PDF of each parameter with its Gaussian fit is shown, as well as the resulting most probable value and 16$^{th}$ and 84$^{th}$ percentiles, which are also marked with solid and dashed lines, respectively. It is indicated with a dashed line the median of the distributions. In black at the top right of each plot we show Pearson’s correlation coefficient. High values of this coefficient are highlighted with an intense blue frame in the plot. Finally, the parameters adopted by the MS are marked with a magenta cross and the final values with an orange star.}
        \label{fig:cornerplot_PARSEC}
    \end{figure*}

\begin{figure*}[h!]
        \centering
        \includegraphics[width=\textwidth]{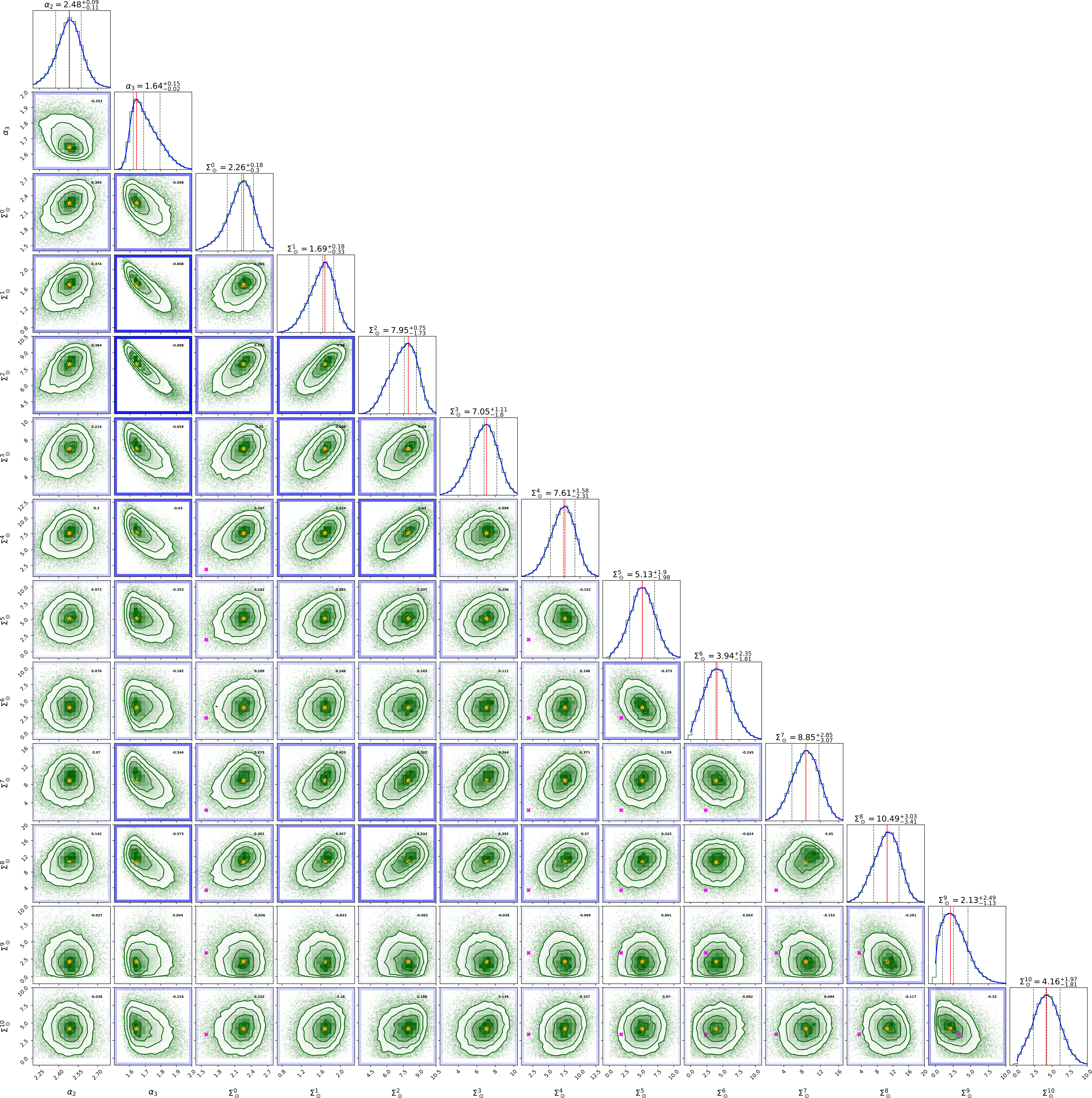}
        \caption{Same as in Fig. \ref{fig:cornerplot_PARSEC} but for execution G13S-13.}
        \label{fig:cornerplot_STAREVOL}
    \end{figure*}

    We show in Figs. \ref{fig:cornerplot_PARSEC} and \ref{fig:cornerplot_STAREVOL} the posterior distributions and the correlations of the BGM FASt inferred parameters for executions G13P-13 (mother simulation built using PARSEC stellar tracks and thick disc SFH not considered in the fit) and G13S-13 (same but using STAREVOL stellar tracks), respectively. The Pearson coefficient $R$, shown in the top-right corner and highlighted in the border of each subplot of the figures, characterizes the linear correlation between two distributions. It can take values from $-1$ (linearly anticorrelated) to 1 (linearly correlated), being 0 the representation of two linearly independent distributions. A description of the most relevant correlations among parameters is given in Sect. \ref{subsect:impact_correlations}.
    

    \section{Colour magnitude diagrams of the fiducial executions} \label{app:complete_cmds}

\begin{figure*}[h!]
        \centering
        \includegraphics[width=\textwidth]{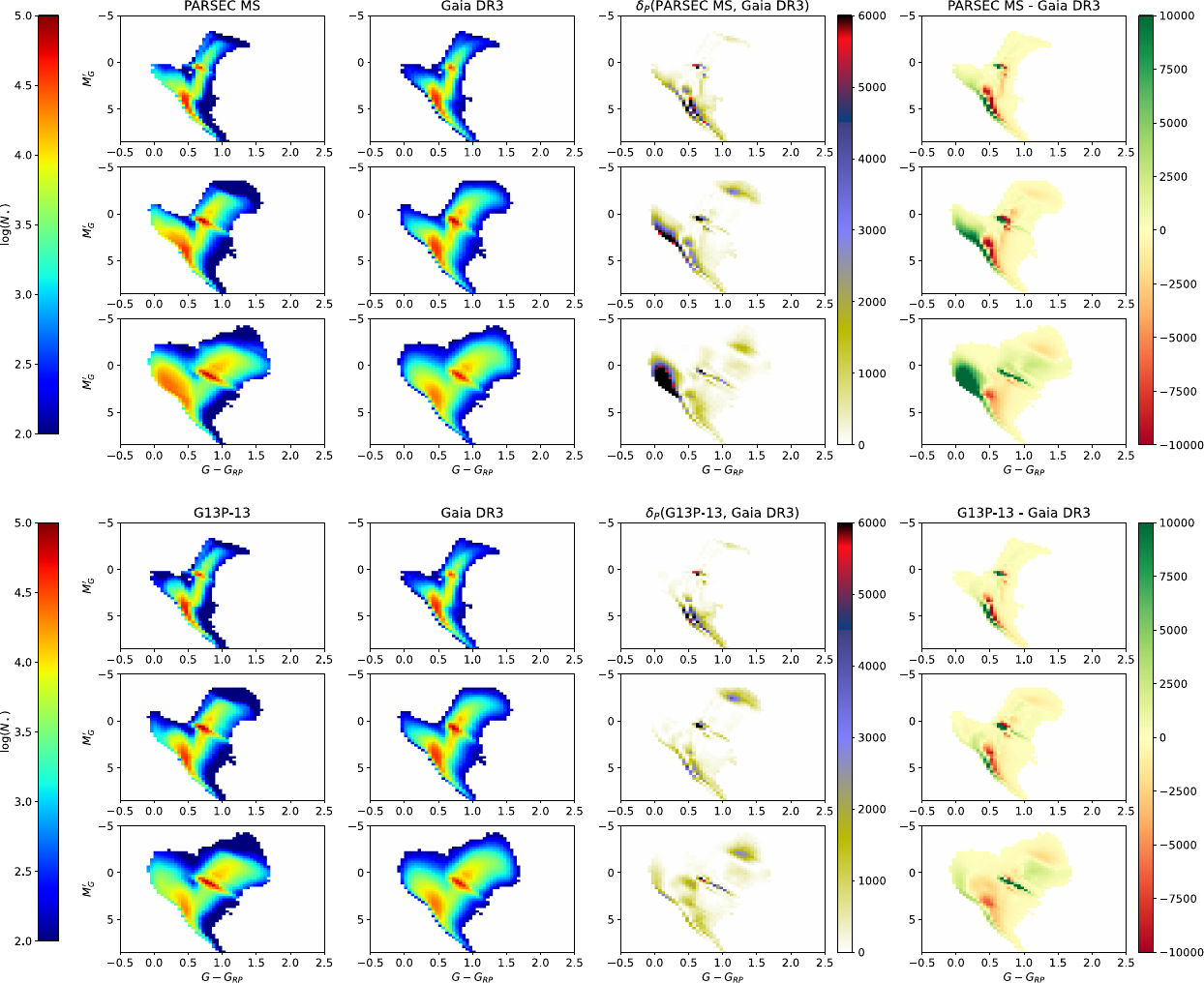}
        \caption{\textbf{Top:} From left to right-hand side, columns show CMDs of the PARSEC MS used in execution G13P-13, {\it Gaia} DR3 $G<13$ data, distance per pixel between them (each of the elements of the summation in Eq. (\ref{eq:distance})), and their absolute difference in number of stars. From the top to the bottom rows, we find the CMDs for the three considered latitude ranges $30<|b|<90$, $10<|b|<30$, $0<|b|<10$. \textbf{Bottom:} same as on top but for the final BGM FASt PS obtained from the application of the fitting process described in Sect. \ref{subsect:basics_BGMFASt} on PARSEC MS.}
        \label{fig:CMD_PARSEC}
    \end{figure*}

\begin{figure*}[h!]
        \centering
        \includegraphics[width=\textwidth]{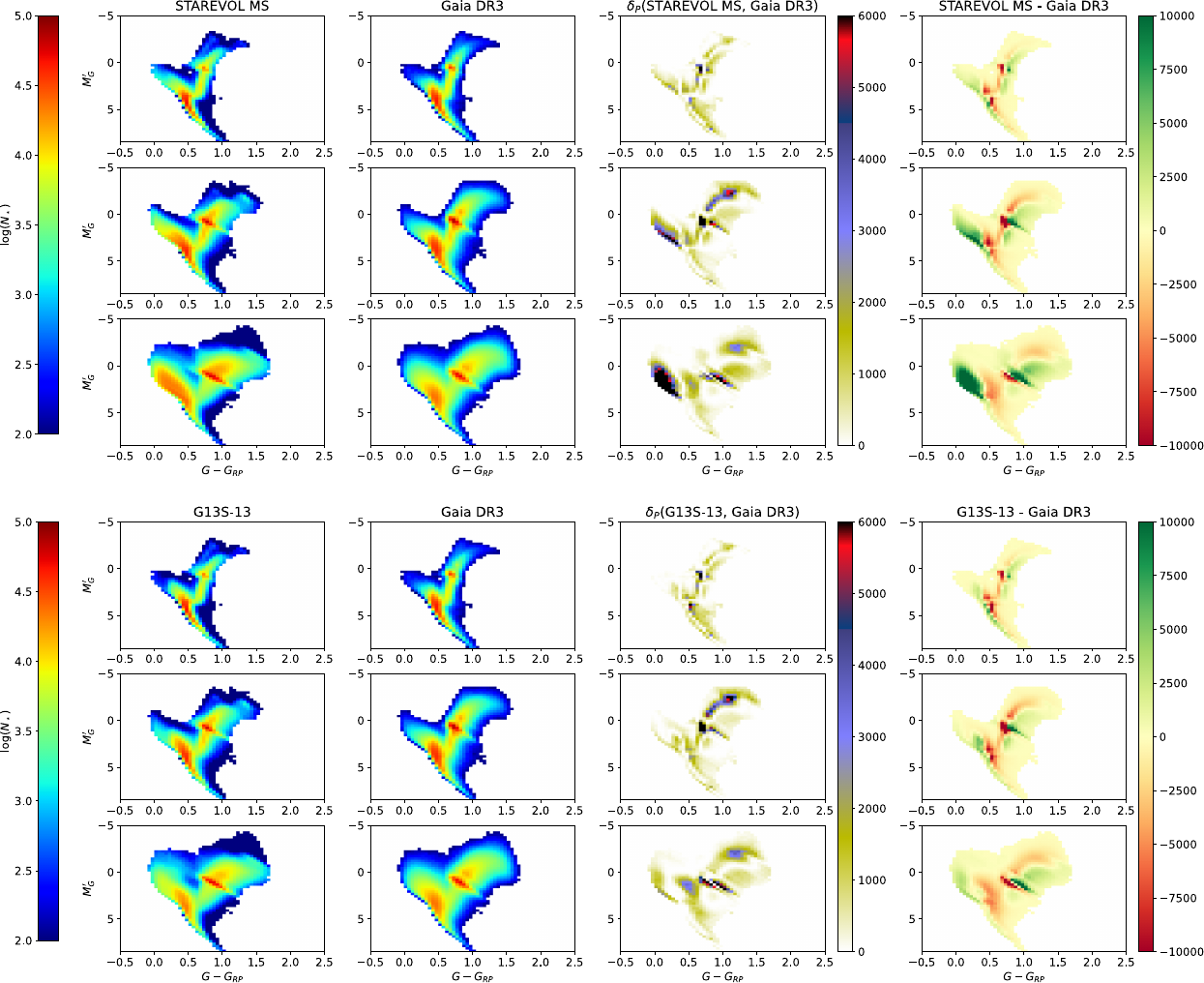}
        \caption{Same as in Fig. \ref{fig:CMD_PARSEC} but for the STAREVOL MS used in execution G13S-13.}
        \label{fig:CMD_STAREVOL}
    \end{figure*}

    Figs. \ref{fig:CMD_PARSEC} and \ref{fig:CMD_STAREVOL} expand the information provided in Fig. \ref{fig:CMD_zoom} and show the complete picture of the comparisons between {\it Gaia} and our model before and after the BGM FASt inference for executions G13P-13 (mother simulation built using PARSEC stellar tracks and thick disc SFH not considered in the fit) and G13S-13 (same but using STAREVOL stellar tracks), respectively. In Table \ref{tab:numbers_and_distances} we present the quantification of the comparisons shown in the aforementioned figures, with the evolution of the distance and the difference in number of stars between {\it Gaia} and the model in terms of the latitude bin. We complement here Sect. \ref{subsect:cmds_evolution} with additional comments analyzing this table.

    In the first column of the table and for each dataset (G13P-13 and G13S-13 before and after BGM FASt, and {\it Gaia} $G<13$ data), we show the total number of stars in the sample and the contribution of each latitude bin. The initial guess in the mother simulations has in all cases too many stars, and the distribution shows an excess of them at low latitudes compared to {\it Gaia}. This is completely solved by the new SFH and IMF. Especially for G13P-13, the discrepancy in the total number of stars with respect to {\it Gaia} data improves from an initial relative contribution of 17\% to a residual value of 2\%, with an almost perfect distribution in the different latitude bins. This is probably mainly due to the significant reduction of stars in the upper main sequence at $|b|<10$ thanks to the new SFH and IMF pointed out in Sect. \ref{subsect:cmds_evolution}. Similar conclusions can be made regarding G13S-13. However, while the distribution between latitude bins clearly improves, the total number of stars in the final PS still falls far from the value in {\it Gaia}, with a relative discrepancy of $\sim$ 8\% mainly concentrated in mid-latitudes ($10<|b|<30$). 

    In the third column of each dataset in Table \ref{tab:numbers_and_distances} we find the total distance between the model and {\it Gaia} before and after the inference and the contribution to this value of the stars in the different latitude bins. Taking the results for G13P-13, we see that the contribution of stars at $|b|<10$ diminishes 10 points throughout the BGM FASt process, from 48.9\% to 38.7\%, while the contribution to the total discrepancy at mid and especially high latitudes increases significantly. This is in concordance with the caveats of this work, explained in detail in Sects. \ref{subsect:limits_and_caveats} and \ref{sect:conclusions}. Our fit up to magnitude $G=13$ lets us mainly tackle issues with young ($\tau<2-3$ Gyr) and intermediate-mass stars, while the characterization of old and low-mass stars remains unsolved. The former population has typically more circular orbits and less vertical heating, which locates their stars in the very thin disc, extensively observed at low latitudes. On the other hand, the old star population is found further away from the Galactic plane and being part of both the thin and the thick discs. That is the reason why their presence is more noticeable at high latitudes. Therefore, the limitations of this work and the nature of the star populations explain why the contribution to the total distance at high latitudes increases significantly with the inference process. 
    
    \begin{table*}[hb!]
    \caption{\textbf{Left:} Number of stars ($N$), absolute difference in star counts ($\Delta N$) and distance defined in Eq. \ref{eq:distance} ($\delta_P$) with respect to {\it Gaia} for both the mother simulations (MSs) and the final BGM FASt pseudo-simulations (PSs), classified by latitude for G13P-13 and G13S-13 executions. We recall that these executions are performed fixing the thick disc and using PARSEC (G13P-13) or STAREVOL (G13S-13) stellar evolutionary models (see Sect. \ref{subsect:parameter_space}). \textbf{Right:} Number of stars ($N$) of the {\it Gaia} $G<13$ sample classified by latitude.} \label{tab:numbers_and_distances}
    \centering
    \begin{tabular}{|cc|ccc|ccc|}
    \hline
      &  & \multicolumn{3}{|c|}{\textbf{G13P-13}} & \multicolumn{3}{|c|}{\textbf{G13S-13}}       \\ 
      &  & $N$ & $\Delta N$ ($\cdot 10^6$) & $\delta_P$ & $N$ & $\Delta N$ ($\cdot 10^6$) & $\delta_P$ \\ \hline
    \multirow{4}{*}{\rotatebox{90}{\textbf{MS}}} & All & 8.57$\cdot 10^6$ & +1.31 & 1.77$\cdot 10^6$ & 8.23$\cdot 10^6$ & +0.97 & 1.83$\cdot 10^6$ \\ 
     & $|b|<10$ & 50.6\% & +1.19 & 48.9\% & 49.3\% & +0.92 & 50.1\% \\
     & $10<|b|<30$ & 32.9\% & +0.17 & 31.4\% & 33.4\% & +0.10 & 35.4\% \\
     & $|b|>30$ & 16.5\% & -0.05 & 19.7\% & 17.3\% & -0.05 & 14.5\% \\ \hline
    \multirow{4}{*}{\rotatebox{90}{\textbf{Final PS}}} & All & 7.11$\cdot 10^6$ & -0.15 & 0.92$\cdot 10^6$ & 6.67$\cdot 10^6$ & -0.59 & 1.36$\cdot 10^6$ \\ 
     & $|b|<10$ & 43.3\% & -0.06 & 38.7\% & 44.6\% & -0.16 & 43.9\% \\
     & $10<|b|<30$ & 36.3\% & -0.07 & 33.0\% & 35.3\% & -0.30 & 37.7\% \\
      & $|b|>30$ & 20.4\% & -0.02 & 28.3\% & 20.1\% & -0.13 & 18.4\% \\ 
     \hline              
    \end{tabular}
    \quad
    \begin{tabular}{|c|c|}
        \hline
         & \textbf{\textit{Gaia} $\pmb{G<13}$} \\ 
         & $N$ \\ \hline
        All & $7.27 \cdot 10^6$ \\
         $|b|<10$ & 43.2\% \\
         $10<|b|<30$ & 36.5\% \\
         $|b|>30$ & 20.3\% \\ 
         \hline
    \end{tabular}
    \end{table*}

\end{appendix}

\end{document}